\documentclass[aps, prd, twocolumn, 10pt, superscriptaddress, nofootinbib]{revtex4-2}
\usepackage[dvipsnames]{xcolor}
\usepackage{amsmath}

\usepackage{graphicx}
\usepackage{subfigure}
\usepackage[percent]{overpic}
\usepackage{slashed}
\usepackage{wrapfig}
\usepackage{tabu}
\usepackage{diagbox}
\usepackage{mathrsfs,amsmath,amssymb,amsthm,amsfonts,tikz,graphicx,accents,hyperref, color}
\usepackage{dsfont,epiolmec, latexsym, stmaryrd, comment}
\usepackage{slashed,ccaption}
\usepackage{mathrsfs, calligra, bbm}
\usepackage{leftidx}
\usepackage{import}
\usepackage{multirow}
\usepackage{amsfonts}
\usepackage{pifont}
\usepackage{tabularx}
\usepackage{cancel}
\usepackage[normalem]{ulem}
\usepackage[utf8]{inputenc}
\usetikzlibrary{intersections,calc}
\usepackage{ifthen}
\usepackage{amsmath}
\usepackage{amsthm}

\usepackage{physics}

\hypersetup{ linktoc=all,
    colorlinks, linkcolor={palatinateblue},
    citecolor={brightpink}, urlcolor={amaranth}
}

\graphicspath{{Images/}}

\definecolor{rosy}{RGB}{230,235,252}
\definecolor{myframetitle}{RGB}{90,89,170}
\definecolor{myblocktitle}{RGB}{140,185,249}
\definecolor{mytitle}{RGB}{10,80,26}

\definecolor{darkgreen}{RGB}{27,130,45}
\definecolor{darkblue}{rgb}{0,0,0.3}
\definecolor{darkred}{rgb}{0.7,0,0}

\definecolor{light gray}{RGB}{220,220,220}
\definecolor{dark purple}{RGB}{108,0,217}
\definecolor{pink}{RGB}{190,20,100}
\definecolor{orang}{RGB}{193,63,0}
\definecolor{green}{RGB}{11,98,17}
\definecolor{darkpink}{RGB}{153,0,76}
\definecolor{bluegreen}{RGB}{0,102,102}
\definecolor{greenlagan}{RGB}{0,102,0}
\definecolor{redgreen}{RGB}{102,102,0}
\definecolor{Redgreen}{RGB}{153,76,0}
\definecolor{vividviolet}{rgb}{0.62, 0.0, 1.0}
\definecolor{amaranth}{rgb}{0.9, 0.17, 0.31}
\definecolor{palatinateblue}{rgb}{0.15, 0.23, 0.89}
\definecolor{brightpink}{rgb}{1.0, 0.0, 0.5}
\definecolor{cornflowerblue}{rgb}{0.39, 0.58, 0.93}
\definecolor{deepcarminepink}{rgb}{0.94, 0.19, 0.22}
\definecolor{radicalred}{rgb}{1.0, 0.21, 0.37}
\definecolor{darkmagenta}{rgb}{0.67, 0, 0.67}

\makeatletter 

\begin{document}

\title{Redshift Dependence of $H_0$ Dipole in Pantheon+ Supernovae}

\author{M.H. Jalali-Kanafi}\email{jalali@ipm.ir} 
\affiliation{School of Physics, Institute for research in fundamental sciences (IPM), P.O.Box 19395-5531, Tehran, Iran}
\author{E. \'O Colg\'ain}\email{eoin.ocolgain@atu.ie}  
\affiliation{Atlantic Technological University, Ash Lane, Sligo F91 YW50, Ireland}
\author{S. Pourojaghi}\email{pourojaghi@ipm.ir}  
\affiliation{School of Physics, Institute for research in fundamental sciences (IPM), P.O.Box 19395-5531, Tehran, Iran}
\author{M.M. Sheikh-Jabbari}\email{jabbari@theory.ipm.ac.ir} 
\affiliation{School of Physics, Institute for research in fundamental sciences (IPM), P.O.Box 19395-5531, Tehran, Iran}

\begin{abstract}
We examine the dipole structure in the local cosmic expansion rate $H_0$ using the Pantheon+ compilation of Type Ia supernovae. We reconstruct directional maps of $H_0$ across the sky within the low-redshift regime, $z_{\rm min} < z < 0.2$, evaluated in the CMB rest frame. To perform a tomographic analysis, we choose 10 different values for $z_{\rm min}\in [0.015 , 0.045]$ range. We find a dipole amplitude $A_{\rm dip}=1.16\pm0.28\ {\rm km/s/Mpc}$ for the lowest bin, monotonically decreasing to $A_{\rm dip} =0.35\pm0.52\ {\rm km/s/Mpc}$, as we increase $z_{\rm min}$. {Since $A_{\rm dip}$ is positive-definite, by construction $A_{\rm dip} > 0$ even in an isotropic universe, thus to assess the statistical significance, we compare to $A_{\rm dip}$ generated from 1000 Monte Carlo simulations based on the $\Lambda$CDM model and the same redshift ranges.} Our results reveal a $2-3\sigma$ dipole pattern (depending on how the significance level is computed) for $z_{\rm min}\lesssim 0.032$. For redshift thresholds where the signal remains statistically significant ($>2\sigma$), the inferred dipole direction points close {to Shapley supercluster and CMB dipole directions}. As $z_{\rm min}$ increases, the dipole amplitude diminishes in line with expectations of a higher redshift isotropic Universe, {but the difference in maximal antipodal $H_0$ determinations, $\Delta\text{H}_0^{\rm max}$, retain a signal of an anisotropy that disappears in the dipole ansatz. The decreasing statistical significance of all estimators with increasing $z_{\rm min}$ suggests that any $H_0$ dipole is a low-redshift feature.}
\end{abstract}

\maketitle

\section{Introduction} \label{sec:Introduction}
\vspace*{-3mm}

The Cosmological Principle (CP), namely the assumption that background cosmology at large enough distances is homogeneous and isotropic and that fluctuations/perturbations around the background are statistically homogeneous and isotropic, {alongside General Relativity}, constitutes the backbone of our current understanding of the universe. The CP in cosmological models is built in through the FLRW background and is made manifest by assuming that the expansion rate  at cosmological scales, the Hubble parameter $H(z)$, is only a function of the redshift $z$ and not the direction over the celestial sphere. 

While the $z$-dependence of $H(z)$ is determined once we choose a specific cosmological model, dealing with low enough redshifts, one can Taylor expand $H(z)$ around $z=0$ and treat it in a model-independent way:
\begin{equation}\label{H-low-z-Taylor}
    H(z)=H_0\left(1+ (1+q_0)z+{\cal O}(z^2)\right)
\end{equation}
As Allan Sandage famously put it in 1970 \cite{Sandage-1970}, {understandably before cosmic microwave background (CMB) anisotropies were fully elucidated},  ``cosmology is a search for two numbers'', the Hubble constant $H_0$ and deceleration parameter $q_0$. 

However, there is mounting evidence that the reality in the sky may not be as simple as depicted under the CP \cite{Aluri:2022hzs}. Breakdown of isotropy, which is less controversial and easier to probe than homogeneity (we have only one vantage point on the universe),  is usually probed through a dipole component in cosmological observables. In this work we will search for a dipole component in $H_0$ in low-$z$, $z<0.2$, within the Pantheon+ supernovae (SNe) dataset \citep{brout2022pantheon+, Scolnic:2021amr}. Since we are working in a low-$z$ regime 
we can either employ  \eqref{H-low-z-Taylor} or use any equivalent two-parameter model, here the flat $\Lambda$CDM model, and provided one fits the parameters, one expects results independent of the cosmological model.

Of course, we are not the first group to search for a dipole in $H_0$. In 2007 McClure \& Dyer \cite{McClure:2007vv} observed arguably the earliest $H_0$ dipole corresponding to $\Delta H_0 \sim 9$ km/s/Mpc in HST KEY project data based on different distance indicators. The KEY project quotes $H_0 = 72 \pm 8$ km/s/Mpc \cite{HST:2000azd}; so, the variation of $H_0$ over the sky is of the same order as the reported error, and may be safely absorbed. In the same year, Schwarz \& Weinhorst \cite{Schwarz:2007wf} identified a $H_0$ dipole in exclusively SNe data through hemisphere decomposition at $z < 0.2$. In terms of directions, \cite{McClure:2007vv} finds a maximum $\Delta H_0$ perpendicular to the CMB dipole direction while  \cite{Schwarz:2007wf} recovers further distinct directions. 

As Type Ia SNe samples improved in size and calibration, the $H_0$ dipole signal has persisted \cite{Bahr-Kalus:2012yjc, Bengaly:2015dza, Javanmardi:2015sfa, Krishnan:2021jmh, Kalbouneh:2022tfw, McConville:2023xav, Perivolaropoulos:2023tdt, Hu:2023eyf,  Lopes:2024vfz, Sah:2024csa} against a backdrop of more general claims that SNe samples are anisotropic \cite{Antoniou:2010gw, Cai:2011xs, Cai:2013lja, Mariano:2012wx, Appleby:2014kea, Zhao:2019azy, Sorrenti:2022zat, Hu:2024qnx, Verma:2024lex, Lopez-Hernandez:2026icv, Zhao:2026}. It is imperative that errors are properly propagated through covariance matrices \cite{BeltranJimenez:2014otq}, and given the limits of SNe data, one can still argue that the anisotropies are not statistically significant \cite{Lin:2015rza, Andrade:2017iam, Andrade:2018eta, Deng:2018jrp, Zhai:2022zif, Bengaly:2024ree, Quintana-Estelles:2026xyp}. Beyond the dipole, one may also study higher order multipoles as data quality improves, e. g. the quadrupole \cite{Heinesen:2021azp, Dhawan:2022lze}. While a statistically significant $(>3\sigma)$ SNe anisotropy currently has no conclusive answer, it is clear that $\Delta H_0$ now varies outside of the $\sim 1\%$ errors claimed \cite{Kalbouneh:2022tfw, McConville:2023xav, Hu:2023eyf, Lopes:2024vfz, Sah:2024csa} and coherent (bulk flow) galaxy peculiar motions provide a physical backstory. 

The literature seems to be converging on this point. Starting with the earlier papers \cite{Kashlinsky:2008ut, Watkins:2008hf, Feldman:2009es, Wiltshire:2012uh, Bolejko:2015gmk, Hoffman:2017ako}, there are claims of a  larger than expected statistically significant local bulk flow in both galaxy clusters \cite{Migkas:2020fza, Migkas:2021zdo} and cosmicflows-4 (CF4) data \cite{Watkins:2023rll, Whitford:2023oww}. Both of these can be interpreted as $H_0$ dipoles \cite{Migkas:2020fza, Migkas:2021zdo, Boubel:2024cmh, Salzano:2025ang} with a statistical significance that may be approaching $4 \sigma$. However, see \cite{Stiskalek:2025cjv, Yasin:2026nte} where the statistical significance is questioned. One can also study bulk flows in SNe samples \cite{Sorrenti:2022zat, Sorrenti:2024ugq, Sorrenti:2024ztg, Do:2026ezx}, but the statistics are poorer. In the big picture, bulk flows may preclude an isotropic Universe at distances up to $300 \, h^{-1}$ Mpc \cite{Courtois:2025xcs}. {Whether one views this locally as an anisotropic $H_0$ or an isotropic $H_0$ with a bulk flow, it implies that the Hubble-Lema\^itre Law, a $20^{\rm th}$ century pillar of cosmology, should be revisited and corrected in the $21^{\rm st}$ century.}

Here, we revisit earlier SNe analysis of  the dipole within Pantheon+ compilation by exploiting slightly smaller patches on the sky. The price one pays for working with smaller patches is fewer SNe and thus poorer statistics, where there is a greater danger one runs into noise, but the potential return is greater directional sensitivity that is diluted when one decomposes in hemispheres, e. g. \cite{Schwarz:2007wf}. As explained, we assume the $\Lambda$CDM model but expect results to be model independent due to the low redshift focus. See \cite{Dhawan:2020xmp} where model independence of $H_0$ is explicitly tested in a similar regime.

We analyze the data in a $[z_{\text{min}}, 0.2]$ range for 10 different values of $z_{\text{min}}$ in $0.015$--$0.045$ range and read $ H_0$ for each patch as we move patches over the sky.  In this way we determine the value of the dipole and its direction at each $z_{\text{min}}$. In our analysis we take special care to properly implement  the covariance matrices and to propagate errors among different observables and estimators in our patch-analyses.  To explore the statistical significance of the results, we compare this with the results of isotropically simulated (mock) data based on the $\Lambda$CDM model. Our results shows a significant dipole component {($> 2 \sigma$)} in $H_0$ in the $z_{\text{min}}<0.032$ range pointing to the same region in the sky as the Shapley supercluster {and the CMB dipole}. We also find that the dipole signal decreases as we move to larger $z_{\text{min}}$ {but any shift in direction away from Shapley/CMB dipole directions coincides with a dipole amplitude consistent with zero, so one is simply fitting noise}. Given the diverging statements about $H_0$ dipoles from SNe in the literature, we compare our results with other reported cosmological dipole signatures in the discussion section.

\vspace*{-5mm}
\section{Data Description} \label{sec:Data Description}
\vspace*{-2mm}

\begin{figure*}[ht!]
    \includegraphics[width=\textwidth]{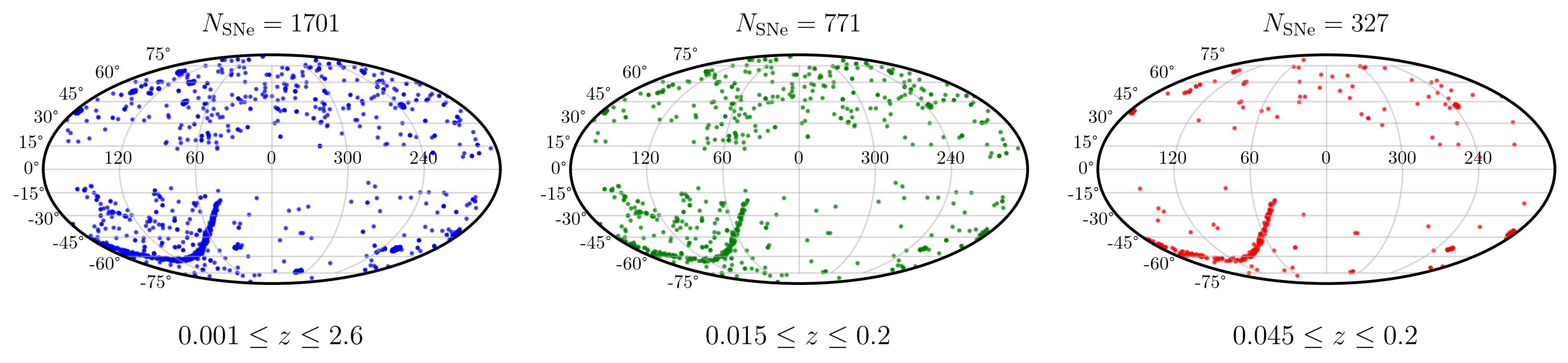}
    \caption{Angular distribution of the Pantheon+ SNe in galactic coordinates for different redshift selections. {As the figures show anisotropies in angular distribution become more prominent as lower redshift SNe are removed.}}
    \label{fig:fig_1}
\end{figure*}

We analyze the Pantheon+ compilation of Type Ia SNe, one of the most systematically calibrated and extensively used datasets currently available for precision cosmology. Pantheon+ \citep{brout2022pantheon+, Scolnic:2021amr}, which supersedes the original Pantheon sample \citep{scolnic2018complete}, includes an expanded SNe sample together with improved photometric calibrations, updated implementations of the SALT2 light-curve framework, and refined treatments of systematic uncertainties. The compilation comprises 1701 spectroscopically confirmed SNe spanning a wide redshift interval, from $z \simeq 0.001$ to $z \simeq 2.26$; 77 of these SNe in the lower $z$ range ($z\leq 0.015$) are in Cepheid host galaxies and are used to calibrate the absolute luminosity $M$. 
In this work, we use the redshifts given in the CMB rest frame, $z_{\rm CMB}$, as provided in the Pantheon+ compilation. These redshifts are corrected for the motion of the observer with respect to the CMB, but do not include additional corrections based on reconstructed peculiar velocity fields. The SNe angular distribution in Galactic coordinates is shown in the left panel of Fig.~\ref{fig:fig_1}.

Here, we restrict our analysis to the low-redshift regime, where anisotropies in the inferred expansion rate are expected to be most sensitive to the impact of nearby large-scale structures and coherent peculiar velocity fields. 
We consider SNe in the redshift interval 
$z_{\rm min} < z < 0.2$,  
where the minimum redshift $z_{\rm min}$ is allowed to vary within the range 
$0.015 \leq z_{\rm min} \leq 0.045$. The upper bound $z=0.2$ is chosen to preserve adequate statistical power while remaining within a redshift range where coherent velocity fields may still imprint measurable directional signatures on the inferred expansion rate. 
At substantially higher redshifts, peculiar velocities become increasingly subdominant relative to the background cosmological expansion, and the angular sampling of SNe becomes progressively sparse, thereby reducing the sensitivity of directional analyses. {Furthermore, at higher redshifts assuming the $\Lambda$CDM model makes results model dependent but one circumvents this concern in the more local Universe.}

In our approach we  systematically vary the lower redshift threshold $z_{\rm min}$. 
We explore ten equally spaced values in the interval 
$0.015 \leq z_{\rm min} \leq 0.045$, 
enabling a redshift tomography of the directional signal. 
The lowest threshold, $z_{\rm min}=0.015$, represents a conservative choice motivated by the need to suppress the impact of highly nonlinear local environments~\citep{lopes2024bulk}. 
At redshifts below $z\sim0.015$, the dominance of galaxy peculiar velocities over the Hubble expansion rate can lead to biases in the local estimation of $H_0$~\citep{riess20162,peterson2022pantheon+}. 

Increasing the threshold to $z_{\rm min}=0.045$ further reduces the influence of very local structures and also alters the imprint of bulk flows on the selected dataset. 
However, adopting substantially larger minimum redshifts results in a strongly anisotropic sky distribution of SNe, which in turn may compromise the robustness of directional analyses. 
For clarity, the sky distribution of SNe for different choices of $z_{\rm min}$ is illustrated in Fig.~\ref{fig:fig_1}, where the middle and right panels correspond to $z_{\rm min}=0.015$ and $z_{\rm min}=0.045$, respectively. 

In the next section, we present the computational pipeline developed to reconstruct directional maps of the local $H_0$ from the SNe sample.

\vspace*{-5mm}
\section{Methodology}\label{sec:Methodology}
\vspace*{-2mm}

The primary objective of this work is to probe possible anisotropies in the local cosmic expansion by reconstructing $H_0$ as a function of sky direction on the celestial sphere. We adopt a directional approach in which the expansion rate is inferred independently over different regions of the celestial sphere, allowing coherent angular patterns in the Hubble flow to be identified and studied in a controlled manner.

\vspace*{-6mm}
\subsection{Directional reconstruction of \texorpdfstring{$H_0$}{H0}}
\label{sec:H0_map}
\vskip -3mm

\begin{figure*}[t]
	\centering
	\includegraphics[scale=0.6]{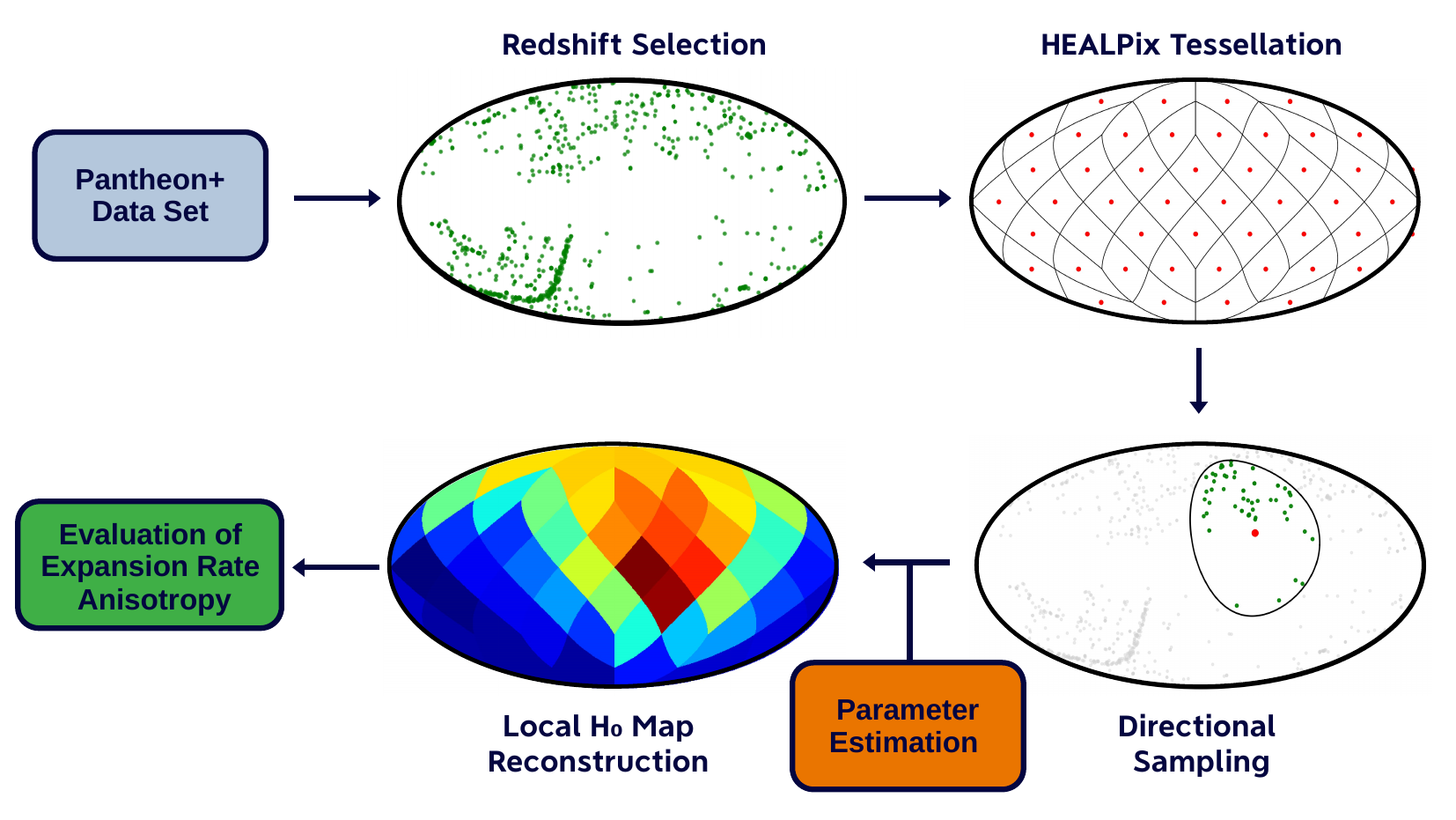} 
    \caption{Schematic illustration of the computational pipeline we adopted to probe anisotropies in the local cosmic expansion. Starting from the Pantheon+ SNe dataset, a redshift selection is applied to define the analysis sample. The sky is  discretized using a HEALPix tessellation and SNe are directionally sampled within overlapping angular caps centered on each pixel. For each sky direction, cosmological parameters are inferred independently within a Bayesian framework, yielding a directional reconstruction of $H_0$. The resulting full‑sky maps form the basis for evaluating anisotropies in the local expansion rate.}
    \label{fig:fig_2}
\end{figure*}

Our reconstruction pipeline, schematically illustrated in Fig.~\ref{fig:fig_2}, is as follows. First, a subsample of Pantheon+ SNe is selected by imposing a specific redshift interval. The sky is then discretized using the HEALPix tessellation scheme \citep{Gorski:2004by} with a low-resolution grid, adopting \( N_{\rm side} = 2 \), such that each pixel center defines a reference direction on the sky. This choice corresponds to a total of \( N = 12 \times N_{\rm side}^2 = 48 \) equal-area pixels uniformly distributed over the celestial sphere. Around each reference direction, we construct a spherical cap with an angular radius of \( 75^\circ \), and the SNe falling within each cap are used to infer local cosmological parameters within a unified theoretical framework. For comparison, a hemisphere has an angular radius of $90^{\circ}$ so the surface area of the spherical cap is approximately $74\%$ of a hemisphere. The resulting estimate of the Hubble parameter is assigned to the corresponding HEALPix pixel,\footnote{The minimum number of SNe in a patch we have  is  29 (for the $z_{\rm min}=0.045$ case). So, we do not need to apply the often-used masking of patches with fewer than \( N_{\rm min} = 20 \) SNe to avoid unstable or poorly constrained parameter estimates.} yielding a discretized map of the locally inferred expansion rate for a given redshift configuration.

{
To infer the local expansion rate in each cap, we model the SNe luminosity--distance relation within a spatially flat $\Lambda$CDM model. For the cosmological parameters $H_0, \Omega_m$, the luminosity distance is defined as
\begin{equation}
\label{eq:eq_1}
{\cal D}_L(z;H_0, \Omega_m) = \frac{ c (1+z)}{H_0}\ 
d_L(z;\Omega_m),
\end{equation}
where
\begin{equation}
d_L(z;\Omega_m)=
\int_0^z 
\frac{dz'}{\sqrt{\Omega_m(1+z')^3 + (1-\Omega_m)}}.
\end{equation}
{For $z<0.2$ and typical Planck value of $\Omega_m\sim 0.3$, $d_L$ computed from the exact $\Lambda$CDM relation or its low-redshift cosmographic Taylor expansion (with deceleration parameter $q_0 = -1+\frac{3}{2} \Omega_m \sim -0.55$) are essentially the same. The sensitivity to $\Omega_m$ is negligible at these low redshifts, thus our analysis is 
effectively model independent.} The distance modulus $\mu$ is defined as the difference between apparent and absolute magnitudes $M$, \( \mu := m - M \), which in terms of \( {\cal D}_L(z) \) becomes,
\begin{equation}
\label{eq:eq_2}
\mu_{\rm th}(z; H_0, \Omega_m) 
= 5 \log_{10}\!\left(\frac{{\cal D}_L(z;H_0, \Omega_m)}{\mathrm{Mpc}}\right) + 25, 
\end{equation} 
where the $\mu_{\rm th} (z)$ subscript reminds us that it depends on the theoretical model.

The total chi-square minimized in each spherical cap is written as
\begin{equation}
\label{eq:chi2tot}
\chi^2_{\rm tot}(\hat n_i)
=
\chi^2_{\rm Cepheid}
+
\chi^2_{\rm SN}(\hat n_i),
\end{equation}
where
\begin{equation}
    \hat{n}_i=(x_i,y_i,z_i), \qquad i=1,\ldots,N,
\end{equation}
denotes the central direction of the $i^{\rm th}$ HEALPix pixel and 
\begin{equation}
\begin{split}
\chi^2_{\rm Cepheid}
=
(\boldsymbol{\Delta}^{\rm Cepheid})^{\rm T}
\mathbf{C}_{\rm Cepheid}^{-1}
\boldsymbol{\Delta}^{\rm Cepheid},\\ 
\chi^2_{\rm SN}(\hat n_i)
=
\left[
\boldsymbol{\Delta}^{\rm SN}(\hat n_i)
\right]^{\rm T}
\left[
\mathbf{C}_{\rm SN}^{\hat n_i}
\right]^{-1}
\boldsymbol{\Delta}^{\rm SN}(\hat n_i),    
\end{split}
\end{equation}
with, 
\begin{equation}
\begin{split}
\Delta_A^{\rm Cepheid}
&=
m_A-M-\mu_A^{\rm Cepheid},\\
\Delta_a^{\rm SN}(\hat n_i)
&=
m_a-M-\mu_{\rm th}(z_a;H_0,\Omega_m).
\end{split}
\end{equation}
{Here, $A=1,\ldots,77$ labels the Cepheid-calibrated SN host galaxies, and $a$ runs over the SNe contained in the angular cap centered on $\hat n_i$, namely $a\in {\rm cap}(\hat n_i)$. The quantities $m_A$, $m_a$, and $\mu_A^{\rm Cepheid}$ are observational inputs, while $\mu_{\rm th}(z_a;H_0,\Omega_m)$ denotes the theoretical distance modulus in flat $\Lambda$CDM, computed from (\ref{eq:eq_1}) and (\ref{eq:eq_2}).}
$\mathbf{C}_{\rm Cepheid}$ is obtained by restricting the full Pantheon+ covariance matrix to the $77$ Cepheid-host SNe and $\mathbf{C}_{\rm SN}^{\hat n_i}$ is obtained by truncating the full Pantheon+ covariance matrix to the Hubble-flow SNe satisfying both the adopted redshift interval $z_{\rm min} < z < 0.2$ and the angular selection defined by the cap centered on $\hat n_i$. 75 of the 77  Cepheid-host SNe are in the redshift range $z<0.015$, {but we never project out these SNe as they serve to calibrate $M$ through $\chi^2_{\rm Cepheid}$}. We note that the wherever we apply redshift or angular cuts, we restrict the Pantheon+ full covariance matrix to the region the cuts are applied and the off-diagonal cross-covariance block between the two subsets is discarded.

Given the truncations, equation \eqref{eq:chi2tot} assumes that any anisotropy only arises from Hubble-flow SNe and not SNe in Cepheid-host galaxies. We minimize $\chi^2_{\rm Cepheid}$ and $ \chi^2_{\rm SN}(\hat n_i)$ simultaneously, assuming that the SN Ia absolute magnitude $M$ is universal across the sky. Involving only the low redshift data, the role of $\chi^2_{\rm Cepheid}$ minimization (over which angular cuts are not applied) is to calibrate $M$. In principle, one could fit the Cepheid-host SNe to $M$ once and replace $\chi^2_{\rm Cepheid}$ in \eqref{eq:chi2tot} with a Gaussian prior on $M$. Our method is essentially similar to the one used in \cite{Malekjani:2023ple}: Use the Cepheid host SNe to calibrate $M$ and then use the rest of SNe to read $H_0, \Omega_m$. Alternatively, one could have divided the Cepheid-host SNe in each angular cap in the sky and get a $M$ which {in principle may vary over the sky if data allows}. Since we are applying $z>0.015$ cuts, together with the angular cuts, our approach is apt to our analyses here. See \cite{McConville:2023xav, Zhai:2022zif} where different  approaches are considered and compared.

For each sky cap, the parameters $(H_0,\Omega_m,M)$ are inferred within a Bayesian framework using uniform priors,
$H_0\in[10,100]$ km\,s$^{-1}$\,Mpc$^{-1}$,
$\Omega_m\in[0.1,0.9]$,
and $M\in[-20,-18]$.
The posterior distribution is sampled with the affine-invariant Markov Chain Monte Carlo sampler \texttt{emcee} \citep{Foreman-Mackey:2012any}. After discarding the burn-in phase and verifying convergence, the median of the posterior distribution of $H_0$ is assigned to the corresponding HEALPix pixel, while its standard deviation is adopted as the associated statistical uncertainty. Repeating this procedure for all sky caps yields a directional reconstruction of the local expansion rate,
$H_0=H_0(\hat n_i;z_{\rm min})$, together with its uncertainty for every redshift configuration considered in this work. These directional $H_0$ maps constitute the input for the dipole reconstruction described in the following section.

\vspace*{-5mm}
\subsection{Dipole in  \texorpdfstring{$H_0$}{H0}, the amplitude, direction and uncertainties}\label{sec:dipole-amp-direc}
\vskip -3mm

We now use the outputs of previous section, namely $H_0=H_0(\hat{n}_i; z_{\min})$ as inputs for the dipole decomposition. Since we are interested in the dipole in the expansion rate, we focus on $H_0$ only, {nevertheless we assure the reader that errors have been propagated to $H_0$ from $M$ and $\Omega_m$}. We use the following expansion ansatz, 
	\begin{equation}
		\begin{aligned}
			{H_0}(\hat{n}_i; z_{\min}) &= H_0^{\mathrm{mono}} + \mathbf{D}\cdot\hat{n}_i \\
			&= H_0^{\mathrm{mono}} + D_x x_i + D_y y_i + D_z z_i ,
		\end{aligned}
		\label{eq:fit_dipole}
	\end{equation}
where $H_0^{\mathrm{mono}}$ denotes the monopole term and $\mathbf{D}=(D_x, D_y, D_z)$ is the dipole vector in Cartesian Galactic coordinates. Later in section \ref{sec:internal-consistency}, we check {the validity of truncating multipoles to a monopole and dipole}. 
The dipole amplitude is then computed from the fitted Cartesian components as:
	\begin{equation}
		A_{\rm dip}=|\mathbf{D}|
		=
		\sqrt{D_x^2+D_y^2+D_z^2}.
	\end{equation}
The dipole direction is obtained by converting the fitted vector $\mathbf{D}$ into longitude and latitude:
	\begin{equation}
		l = \tan^{-1}\big(\frac{D_y}{D_x}\big), \qquad 
		b = \sin^{-1}\left(\frac{D_z}{{|D|}}\right),
		\label{eq:lb}
	\end{equation}
with $l$ and $b$ mapped onto the intervals $0^\circ \leq l < 360^\circ$ and $-90^{\circ} \leq b \leq 90^{\circ}$, respectively.

{ To estimate the four parameters for any given $z_{\min}$ we use a generalized least-squares (GLS) fit to account for both the statistical uncertainties of individual patches and the shared systematic uncertainties across the sky. We have 4 {parameters} in each $z_{\min}$, $H_0^{\mathrm{mono}}, D_x, D_y, D_z$. We note that, as explained above the errors in $\Omega_m$ and $M$ have already been considered in the inferred values of ${H_0}( z_{\min};\hat{n}_i)$ through a fully correlated systematic error across all sky patches. Since our analysis is restricted to $z < 0.2$, the theoretical modulus $\mu_{\text{th}}(z)$ has negligible dependence on $\Omega_m$, so its value is expected  not to affect the rest of parameters in an essential way, while it may impact the errors on ${H_0}(\hat{n}_i; z_{\min})$. Moreover, as discussed,  $M$ is inferred from minimization of $ \chi^2_{\rm Cepheid}$ which is independent of the patches on the sky. 

Let $\boldsymbol{\theta}$ denote the parameter 4-vector to be estimated, $\mathbf{H_0}$ the $N$-data vector containing the measured $H_0$ values in each patch ($N=48$ in our case) and, $\mathbf{X}$ the $4\times N$ design matrix:
		\begin{equation}
			\begin{split}
				\boldsymbol{\theta} &= (H_0^{\mathrm{mono}}, D_x, D_y, D_z)^{\rm T}, \\
				\mathbf{H_0} &= (H_0(\hat{n}_1), \dots, H_0(\hat{n}_N))^{\rm T},\\
				\mathbf{X}_i &= (1, x_i, y_i, z_i), \qquad i=1,\cdots,N.
			\end{split}
		\end{equation}
To fix the monopole and dipole we minimize 
	\begin{equation}\label{Cov-Mat-Sigma-ij}
		\chi^2_{\boldsymbol{\theta}} := (\mathbf{H_0} - \mathbf{X}\boldsymbol{\theta})^{\rm T}\, \boldsymbol{{\Sigma}}^{-1}\, ( \mathbf{H_0} - \mathbf{X} \boldsymbol{\theta})
	\end{equation}
where $\boldsymbol{\Sigma}_{ij}$  is a $N \times N$ covariance matrix,
\begin{equation}\label{Sigma-ij-def}
    \boldsymbol{\Sigma}_{ij}:=\textbf{Cov}(H_0(\hat{n}_i), H_0(\hat{n}_j)). 
\end{equation} 
The $\chi^2_{\boldsymbol{\theta}}$ can be minimized analytically to get the best fit parameters through the expression 
	\begin{equation}
		{\boldsymbol{\theta}}_{\text{best fit}} = (\mathbf{X}^{\rm T} \boldsymbol{\Sigma}^{-1}\ \mathbf{X})^{-1}\ \mathbf{X}^{\rm T} \boldsymbol{\Sigma}^{-1} \mathbf{H_0}.
		\label{eq:theta_gls}
	\end{equation}

What remains is to determine $N \times N$ covariance matrix $\boldsymbol{\Sigma}$ for which we resort to statistically isotropic Monte Carlo realizations,  the \textit{MC-lcdm} simulations,  discussed in the next subsection. 
\vspace*{-6mm}
\subsection{Monte Carlo simulations and the covariance matrix \texorpdfstring{$\boldsymbol{\Sigma}_{ij}$}{Sigma}}\label{Monte Carlo simulations}
\vskip -3mm

The Monte Carlo simulations employed in this work serve two complementary purposes. First, they provide an empirical estimate of the covariance matrix of the reconstructed directional $H_0$ map,  $\boldsymbol{\Sigma}_{ij}$. Second, they establish the distribution of the dipole amplitude expected under the null hypothesis of statistical isotropy, providing an isotropic reference sample, allowing to quantify the significance of the observed anisotropic signal.

\textbf{\textit{MC-lcdm} simulations.} We generate an ensemble of statistically isotropic mock catalogs, denoted as \textit{MC-lcdm}, constructed from a fiducial $\Lambda$CDM cosmology while preserving the observational characteristics of the Pantheon+ sample. For each redshift configuration considered in this work (each $z_{\min}$), we first determine the best-fitting full-sky cosmological parameter triplet $(H_0,\Omega_m,M)$, through minimizing $\chi^2_{\rm Cepheid}+\chi^2_{\rm SN}$, without the angular cuts, {which provide the input values for the mocks.}

For each Hubble-flow SNe in the spherical cap with redshift $z_i$ we retain the angular coordinates and identify apparent magnitudes
$
m_i=\mu_{\rm th}(z_i;H_0,\Omega_m)+M,
$
before generating new $m_i$ values randomly in a Gaussian distribution with input mean $m_i$ and the full Pantheon+ covariance matrix. This procedure preserves the complete covariance structure of the Pantheon+ data, including both statistical and systematic uncertainties together with all correlations between different SNe. Consequently, every mock realization retains the original sky positions, redshift distribution and observational covariance of the survey while remaining statistically consistent with an isotropic $\Lambda$CDM universe.

For each realization, we repeat the complete analysis pipeline described in Sec.~\ref{sec:H0_map}, 
to build each row of a $N_{\rm sim}\times N$ matrix
\begin{equation}\label{HnN}
\mathbf{H}_{\rm sim}
=
\begin{bmatrix}
H^{(1)}_0(\hat n_1) & \cdots & H^{(1)}_0(\hat n_N)\\
\vdots & \ddots & \vdots\\
H^{(N_{ \rm sim})}_0(\hat n_1) & \cdots & H^{(N_{\rm sim})}_0(\hat n_N)
\end{bmatrix},
\end{equation}
where $N_{\rm sim}$ denotes the number of \textit{MC-lcdm} realizations and $N=48$ is the number of HEALPix patches.

\textbf{Covariance matrix $\boldsymbol{\Sigma}_{ij}$.} The covariance matrix  $\boldsymbol{\Sigma}_{ij}$ is defined through  \eqref{Sigma-ij-def} and can be computed using the ensemble of  reconstructed maps \eqref{HnN}. From there one calculates the $1\times N$ vectors of average values of $H_0(\hat n_i)$, $i=1,\ldots,N$, across the $N_{\rm sim}$ simulations, $\overline{H}_0(\hat n_i)$, 
\begin{eqnarray}
\overline{\mathbf{H}}
 &=& [\overline{H}_0(\hat n_1),\ldots,\overline{H}_0(\hat n_N)].
\end{eqnarray}
In terms of which, we reconstruct  the $N\times N$ covariance matrix,
\begin{equation}
\boldsymbol{{\Sigma}} =\frac{1}{N_{\rm sim}-1}\widetilde{\mathbf{H}}^{T}\widetilde{\mathbf{H}}, \qquad  \widetilde{\mathbf{H}} :=\mathbf{H}_{\rm sim}-\mathbf{1}_{N_{\rm sim}}^{T}\!\cdot\!\overline{\mathbf{H}}   
\end{equation}
where  $\mathbf{1}_{N_{\rm sim}} = [1,\ldots,1]$ is an $1\times {N_{\rm sim}}$ vector. The rest of the analyses will follow as outlined before, by reconstructing the dipole  through the GLS estimator described in Sec.~\ref{sec:dipole-amp-direc}, as well as the dipole for the $N_{\rm sim}=1000$ independent realizations.

We close this section with the comment that other mock-construction strategies can also be considered. A detailed comparison of different simulation schemes and their impact on the inferred cosmological dipole will be presented in a companion paper.

\vspace*{-4mm}
\section{Results} \label{sec:Results}
\vspace*{-2mm}
We present our results on the analysis of $H_0(\hat n; z_{\text{min}})$ maps in two subsections. Section \ref{sec:H0-dipole-tomography} is devoted to what is technically a $H_0$ dipole, its amplitude and direction as a function of $z_{\text{min}}$. While it is usually expected that the dipole is the dominant contributor to anisotropy, not all anisotropy features in $H_0(\hat n;z_{\text{min}})$ are captured in the dipole, as {one cannot preclude higher multipoles, e. g. \cite{Heinesen:2021azp, Dhawan:2022lze}. \textit{A priori}, the higher multipoles may be of comparable magnitude or suppressed by orders of magnitude}. In section \ref{sec:internal-consistency} we  study this by focusing maximum directional contrast in $H_0(\hat n; z_{\text{min}})$ maps across opposite sky patches. 

\vspace*{-6mm}
\subsection{Dipole amplitude  and its evolution with \texorpdfstring{$z_{\min}$}{zmin}}\label{sec:H0-dipole-tomography}
\vskip -3mm

\begin{figure*}[ht!]
    \includegraphics[width=\textwidth]{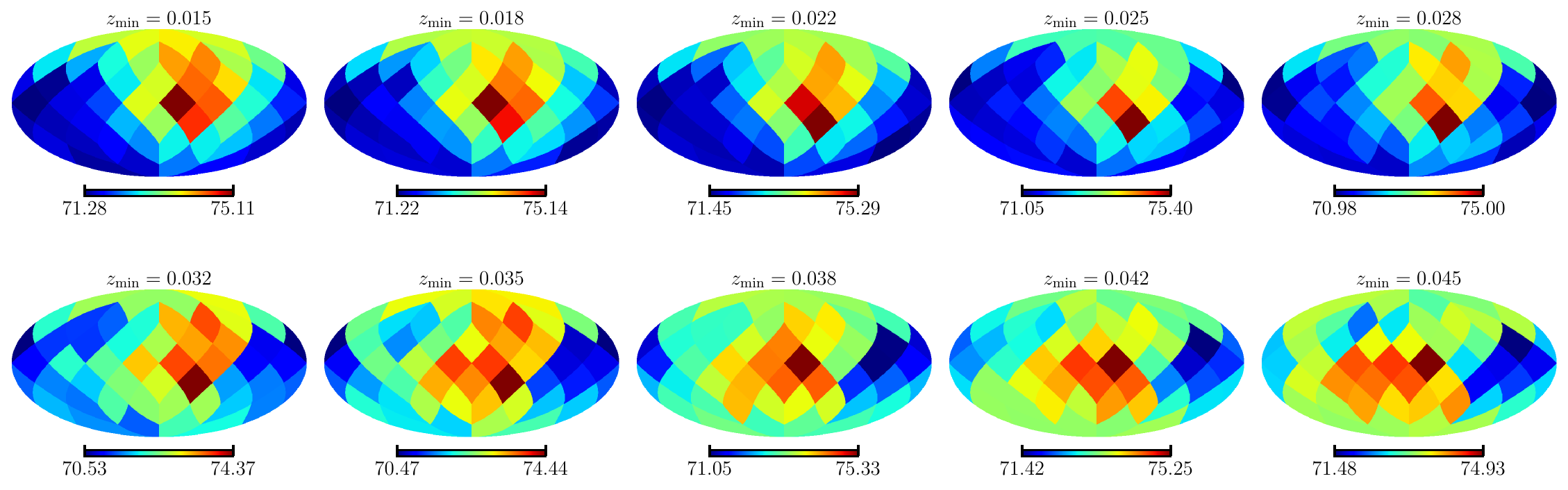}
    \caption{Directional maps of the locally inferred $H_0$ reconstructed from the Pantheon+ sample for different $z_{\text{min}}$, as indicated in each panel. In all cases, the upper redshift limit is fixed to $z_{\max}=0.2$. The maps are obtained using a HEALPix discretization with $N_{\rm side}=2$ and overlapping angular caps of radius $75^\circ$. }
    \label{fig:fig_3}
\end{figure*}
We start by presenting the reconstructed $H_0$ maps for different redshift selections. Fig.~\ref{fig:fig_3} shows the $H_0 (\hat{n})$ maps for 10 choices of the minimum redshift cut $z_{\rm min} \in [0.015, 0.045]$, while the upper limit is fixed to $z_{\max}=0.2$. For the lower values of $z_{\rm min}$, the maps display a clear large-scale angular pattern, consistent with a dipole signal. As the minimum redshift threshold is increased, this pattern becomes progressively less pronounced: the anisotropic signal of the $H_0 (\hat{n}; z_{\rm min})$ maps depends sensitively on the adopted lower redshift cut. These visual trends motivate a quantitative characterization of the revealed dipole signal.

{\begin{table*}[htb!]
	\centering
	\begin{tabular}{|c||  c|c | c|  c|  c| }
		\hline
		$z_{\min}$ &  $H_0^{\text{max}}$ & $H_0^{\text{min}}$ &  ${A_{\rm dip}}$  & $p$-value & $\begin{array}{c} \text{Dipole significance}\\ \text{estimator}\ S \ \eqref{eq:significance}\end{array}$  \\
		\hline
		0.015 &  $75.11 \pm 1.27$ & $71.28 \pm 1.07$ & $1.16 \pm 0.28$  & $ < 0.001 \; (>3.1\sigma)$ & $2.5\sigma$ \\
		0.018 & $75.14 \pm 1.24$ & $71.22 \pm 1.11$ & $1.01 \pm 0.28$  & $ < 0.001 \; (>3.1\sigma)$ & $1.9\sigma$ \\
		0.022 &  $75.29 \pm 1.43$ & $71.45 \pm 1.15$ & $0.87 \pm 0.30$ &   $ 0.01299 \; (2.2\sigma)$  & $1.4\sigma$\\
		0.025 &  $75.40 \pm 1.41$ & $71.05 \pm 1.20$ & $0.91 \pm 0.34$  &  $ 0.01698 \; (2.1\sigma)$ & $1.2\sigma$ \\
		0.028 & $74.99 \pm 1.51$ & $70.98 \pm 1.23$ & $0.93 \pm 0.39$  &  $ 0.04196\: (1.7\sigma)$ & $1.0\sigma$\\
		0.032 & $74.36 \pm 1.59$ & $70.53 \pm 1.27$ & $1.05 \pm 0.39$  &  $ 0.01698 \; (2.1\sigma)$ & $1.2\sigma$\\
		0.035  & $74.44 \pm 1.66$ & $70.47 \pm 1.31$ & $0.87 \pm 0.43$  &   $ 0.1139 \; (1.2\sigma)$  & $0.6\sigma$\\
		0.038 & $75.33 \pm 1.54$ & $71.05 \pm 1.32$ & $0.56 \pm 0.46$  &   $ 0.4725 \; (0.1\sigma)$ & $0.0\sigma$\\
		0.042 &  $75.25 \pm 1.61$ & $71.42 \pm 1.41$ & $0.49 \pm 0.45$  &  $ 0.5784 \; $ ($-0.2 \sigma$) & $- 0.2\sigma$\\
		0.045 & $74.93 \pm 1.71$ & $71.48 \pm 1.45$ & $0.35 \pm 0.52$  &  $0.8452 \; $ ($-1.0 \sigma$) & $ - 0.5\sigma$ \\
		\hline
	\end{tabular}
	\caption{Best-fit for maximum and minimum value of $H_0$ at each $z_{\min}$ (cf. Fig. \ref{fig:fig_3}), and dipole parameters, the amplitude,  the dipole significance and $p$-value computed using the information in Fig.~\ref{fig:fig_5} and the dipole significance estimator computed as in \eqref{eq:significance}. As explained in the main text, we crucially note that one cannot use the error on $A_{\text{dip}}$ in the fourth column to assess the $\sigma$-level significance of the dipole; our isotropic null hypothesis is not the naive $A_{\text{dip}}=0$, but is based on the \textit{MC-lcdm} realizations. {Note also that the $p$-value is more realistic, while $S$ is a more conservative estimator that propagates an additional error.}}\label{tab:dipole-amplitude}
\end{table*}}

To quantify this behavior, we extract the dipole component—both amplitude and direction—of the $H_0(\hat{n}; z_{\rm min})$ maps using equations in section \ref{sec:dipole-amp-direc}. The resulting dipole amplitude as a function of $z_{\rm min}$ is shown in Fig.~\ref{fig:fig_41}. The discussions on the direction of the dipole and its evolution  with $z_{\rm min}$ will be discussed in section \ref{sec:direction}. As can be seen in Table \ref{tab:dipole-amplitude}, the dipole amplitude systematically decreases as $z_{\rm min}$ increases, falling from {a non-zero value at $4.1 \sigma$}
\begin{equation}\label{A-dip-lowest-z}
    A_{\rm dip}=1.16 \pm0.28~{\rm km\,s^{-1}\,Mpc^{-1}} \ \text{at} \ z_{\rm min}=0.015
\end{equation} 
to {a value consistent with zero within $1 \sigma$}
\begin{equation}\label{A-dip-highest-z}
  A_{\rm dip}=0.35  \pm 0.52~{\rm km\,s^{-1}\,Mpc^{-1}} \ \text{at} \ z_{\rm min}=0.045.   
\end{equation}
This trend indicates that the strength of the dipole anisotropy signal drops as closer SNe are excluded from the analysis; the dipole amplitude drops as we move to higher $z_{\text{min}}$. This has been shown in  Fig.~\ref{fig:fig_41}. 
\begin{figure}[ht!]
    \centering
    \includegraphics[width=\linewidth]{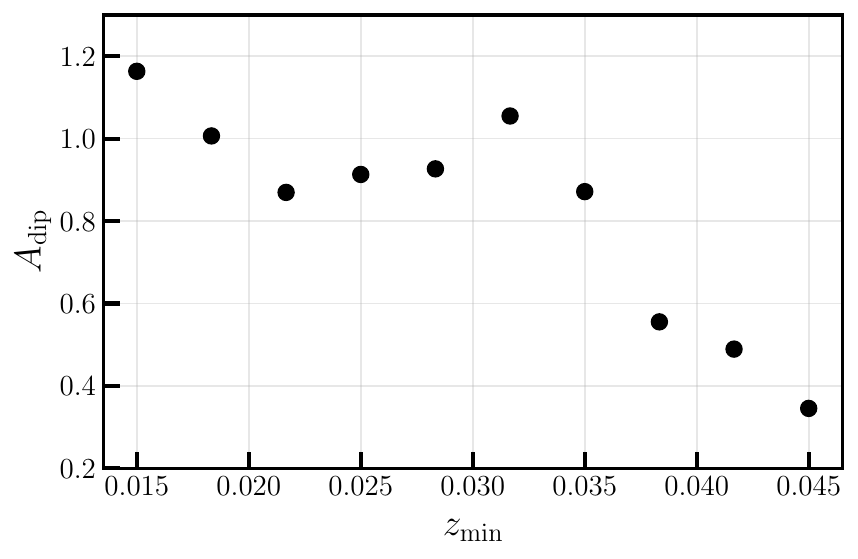}
    \caption{Dipole amplitude best-fit values as a function of  $z_{\rm min}$. Note that the error-bar and the $\sigma$-level significance of the dipole should be assessed using an appropriately constructed null hypothesis. Two such assessments are given in Fig.~\ref{fig:fig_6}.}
    \label{fig:fig_41}
\end{figure}

To assess the statistical significance of the observed anisotropic signal, we should compare the dipole amplitude inferred from the real data in the fourth column of Table \ref{tab:dipole-amplitude}, with an appropriately constructed null hypothesis, namely isotropically constructed reference configurations. {Eq.~\eqref{A-dip-lowest-z} simply shows that $A_{\rm dip}$ is non-zero but does not establish its statistical significance  in a setting where the expectation value is non-zero $\langle A_{\rm dip} \rangle > 0$ because $A_{\rm dip}$ is positive definite.} For the null hypothesis we use isotropic \textit{MC-lcdm} realizations described in section \ref{Monte Carlo simulations}, which by construction preserve the observational characteristics of the Pantheon+ sample.} {As a final comment on Table \ref{tab:dipole-amplitude}, we quote $H_0^{\rm max}$ and $H_0^{\rm min}$ values. In principle, one could attempt to assign a statistical significance to the difference through the errors, but these values need not be from antipodal (opposite) spherical caps, so this may not correspond to a dipole signal and moreover, they can share SNe and are unlikely to be statistically independent. We address this later in Table \ref{tab:deltaH}.}

In Fig.~\ref{fig:fig_5}, each panel presents the distribution of dipole amplitudes derived from $1000$ \textit{MC-lcdm} realizations (shown as blue histograms). The vertical red line indicates the dipole amplitude measured from the real data, while the shaded red band denotes its corresponding $1\sigma$ uncertainty for each choice of $z_{\rm min}$. As seen in Fig.~\ref{fig:fig_5}, the dipole amplitude measured from the real data exhibits the largest deviation from the distribution of \textit{MC-lcdm} realizations at the lowest redshift threshold, $z_{\rm min}=0.015$. As $z_{\rm min}$ increases, this deviation gradually decreases, with the red band moving towards the tail of the simulated distribution around $z_{\rm min}=0.028$, and eventually approaching its center for $z_{\rm min}=0.045$. This information has been used to quantify the dipole $p$-value ($\sigma$-level significance)  column in Table \ref{tab:dipole-amplitude}, as well as the red squares in Fig. \ref{fig:fig_6}. 
{The $p$-value is then computed as
\begin{equation}
p=\frac{N_{\rm cross}+1}{N_{\rm sim}+1},
\end{equation}
where $N_{\rm cross}$ is the number of mocks crossing the red line in Fig.~\ref{fig:fig_5} and $N_{\rm sim}$ is the total number of mock realizations. The addition of one to both the numerator and denominator corresponds to the standard finite-sample correction for Monte Carlo $p$-value estimation, ensuring that the smallest attainable $p$-value is $1/(N_{\rm sim}+1)$ rather than zero. Hence, for $N_{\rm sim}=1000$, the maximum reportable significance is approximately {$3.1 \sigma$}.}

\begin{figure*}[t]
    \includegraphics[width=\textwidth]{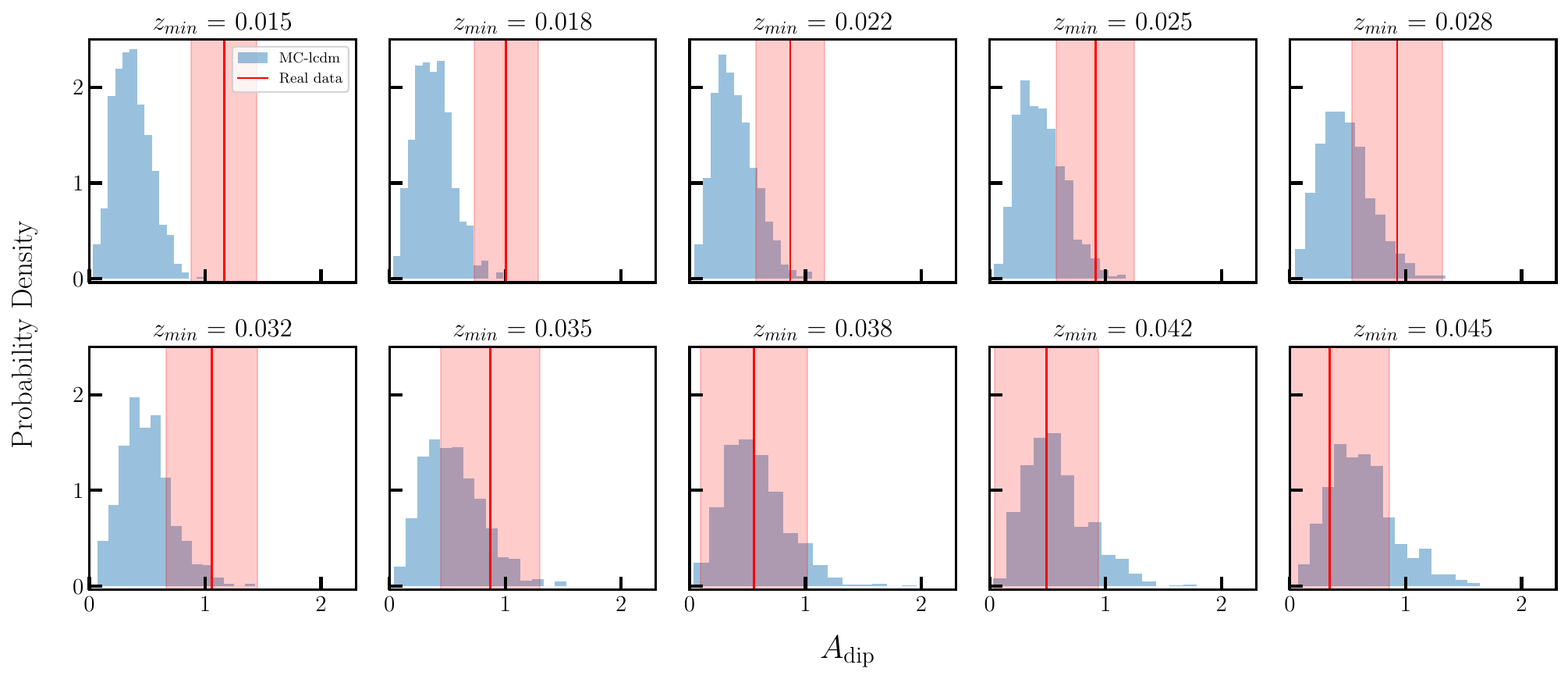}
    \caption{Distribution of the $H_0(\hat{n})$ map dipole amplitudes obtained from $1000$ \textit{MC-lcdm} realizations for different choices of the minimum redshift $z_{\rm min}$ (blue histograms). The vertical red line and the band around it is read from the fourth column in Table \ref{tab:dipole-amplitude}. 
    }
    \label{fig:fig_5}
\end{figure*}

As another more conservative measure for the statistical significance of the observed dipole signal, one can  define the significance estimator
\begin{equation}\label{eq:significance}
S =
\frac{{ A_{\rm dip}^{\rm obs}-\langle A_{\rm dip}^{\rm MC}\rangle }}
{\sqrt{\sigma_{\rm obs}^{2}+\sigma_{\rm MC}^{2}}},
\end{equation}
where $A_{\rm dip}^{\rm obs}$ is the dipole amplitude measured from the observational data, $\langle A_{\rm dip}^{\rm MC}\rangle$ is the mean dipole amplitude obtained from the \textit{MC-lcdm} realizations, $\sigma_{\rm obs}$ denotes the uncertainty of the observed dipole amplitude, and $\sigma_{\rm MC}$ is the standard deviation of the dipole amplitudes derived from the \textit{MC-lcdm} ensemble. The values for this estimator are given in the last column of Table \ref{tab:dipole-amplitude}.   
$S$ has an overall decreasing trend with increasing $z_{\rm min}$, with a small bump around $z_{\rm min}=0.03$, {which reflects a similar sharp decrease in $p$-values at the same $z_{\rm min}$}. In Fig.~\ref{fig:fig_6} we have plotted the two dipole significance levels (the last two columns of Table \ref{tab:dipole-amplitude}). As we see, depending which $\sigma$-level measure is used, the dipole signal in the lowest $z_{\min}$ is {$2.5-3.1 \sigma$}. {As is clear from Fig. \ref{fig:fig_5}, the red band denoting $A_{\rm dip}$ and its $1 \sigma$ confidence interval essentially coincides with $A_{\rm dip} =0$ beyond $z_{\rm min} = 0.038$, so at this point the dipole signal effectively disappears and one is into noise. For context, the Shapley supercluster has an estimated extent of $0.03 \lesssim z \lesssim 0.06$ \cite{Quintana:2000vb}, so the disappearance of the dipole coincides with the beginning of Shapley. While  suggestive, it reinforces the difficulties trying to map out the effect of Shapley in Pantheon+ SNe.}

To facilitate comparison between $S$ $\sigma$ values, we have converted the $p$-values into $\sigma$ in brackets, cf. Table \ref{tab:dipole-amplitude} and Fig.~\ref{fig:fig_6}. While it is reassuring that $S$ and $p$-value estimators both exhibit a similar $z_{\min}$ behavior, since $S$ propagates an additional error $\sigma_{\rm obs}$ for Gaussian mock distributions we expect $p$-values to lead to larger absolute $\sigma$ values, cf. Fig.~\ref{fig:fig_6}. Modulo a small discrepancy that can easily arise from non-Gaussian tails in the direction of larger $A_{\rm dip}$ values, we find that this is the case.
\begin{figure}[ht]
    \centering
    \includegraphics[width=\linewidth]{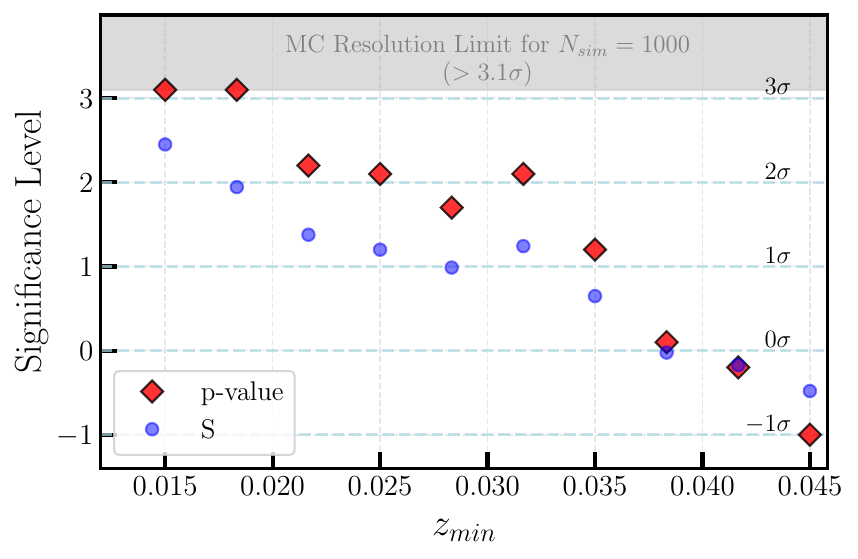}
    \caption{Statistical significances of the dipole amplitude as a function of  $z_{\rm min}$. We have two $\sigma$-level significance estimators. Both are based on the \textit{MC-lcdm} realizations as null hypothesis and may be computed using the information in Fig.~\ref{fig:fig_5}. The more realistic one, in the plot depicted in red,  compares the values of the \textit{MC-lcdm} realizations with the best-fit value of $A_{\text{dip}}$ in Table \ref{tab:dipole-amplitude}. The more conservative one $S$ (\ref{eq:significance}), in the plot depicted in blue, takes into account the errors in the quoted $A_{\text{dip}}$ in Table \ref{tab:dipole-amplitude}. These are respectively plotting the last two columns in Table \ref{tab:dipole-amplitude}. As we see both indicators follow the same pattern with the redshift $z_{\rm min}$.}
    \label{fig:fig_6}
\end{figure}

\vspace*{-6mm}
\subsection{On the significance of covariance matrix}\label{sec:cov-mat}
\vskip -3mm

One of the crucial points in  the parameter estimations in data fitting is appropriately determining and utilizing the covariant matrix. In our problem we start with the covariance matrix of Pantheon+ compilation. To obtain 
$H_0(\hat{n};z_{\min})$ maps, we cut the Pantheon+ covariance matrix over each patch around a given $\hat{n}$ direction on the sky. Then, as discussed in section \ref{sec:Methodology}, we fit the dipole model/ansatz \eqref{eq:fit_dipole} into $H_0(\hat{n};z_{\min})$ maps. We hence need to construct a covariance matrix over the $N\times N$ space of patches, in which the covariance among different patches should also be considered. Moreover, due to dealing with low redshift ($z< 0.2$) SNe,  $\Omega_m (\hat{n};z_{\min})$ values are not expected to shift the best fits values for $H_0(\hat{n};z_{\min})$ maps, they can impact the errors on expansion rate dipole estimations. All in all, it is crucial to consider full covariance matrix to propagate all the errors correctly and avoid underestimating the errors. 

As discussed in section \ref{Monte Carlo simulations}, the covariance matrix we construct using the MCMC simulations is capable of capturing the full covariance matrix, directly inferred from the Pantheon+ covariance matrix. For the comparison and a crude estimate of full covariance matrix, one may naively take a diagonal $N\times N$ covariance matrix, essentially considering the Pantheon+ covariance matrix cut over each patch, and take patches as independent. In Fig.~\ref{fig:fig_8}  we have shown how much one can underestimate the error on dipole amplitude $A_{\text{dip}}$ and the conservative dipole significance estimator $S$ \eqref{eq:significance},  if one uses the diagonal covariance matrix instead of the full covariance matrix (that we have used).

\begin{figure*}[t]
\centering
\includegraphics[width=0.49\textwidth]{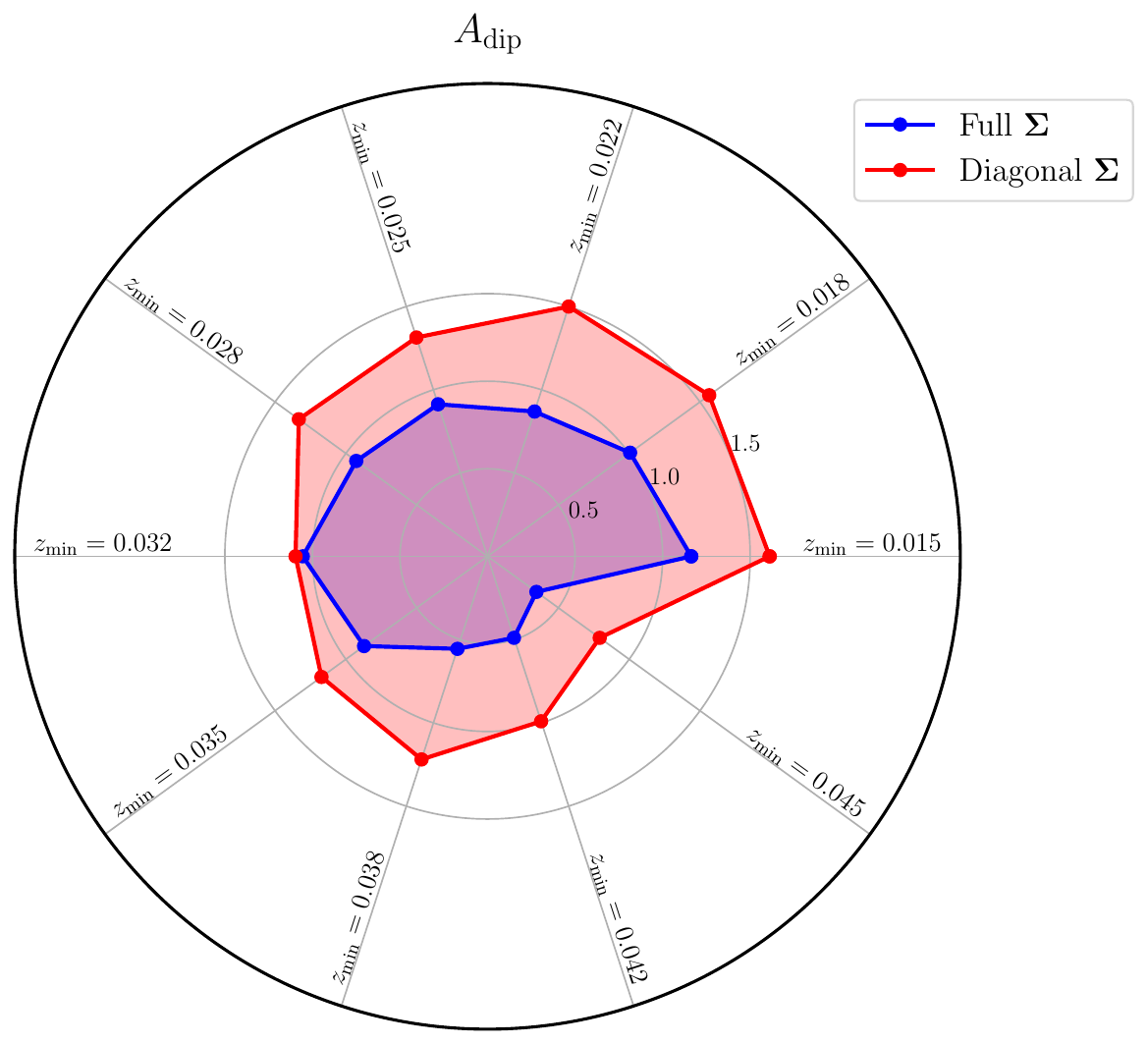}\hfill
\includegraphics[width=0.49\textwidth]{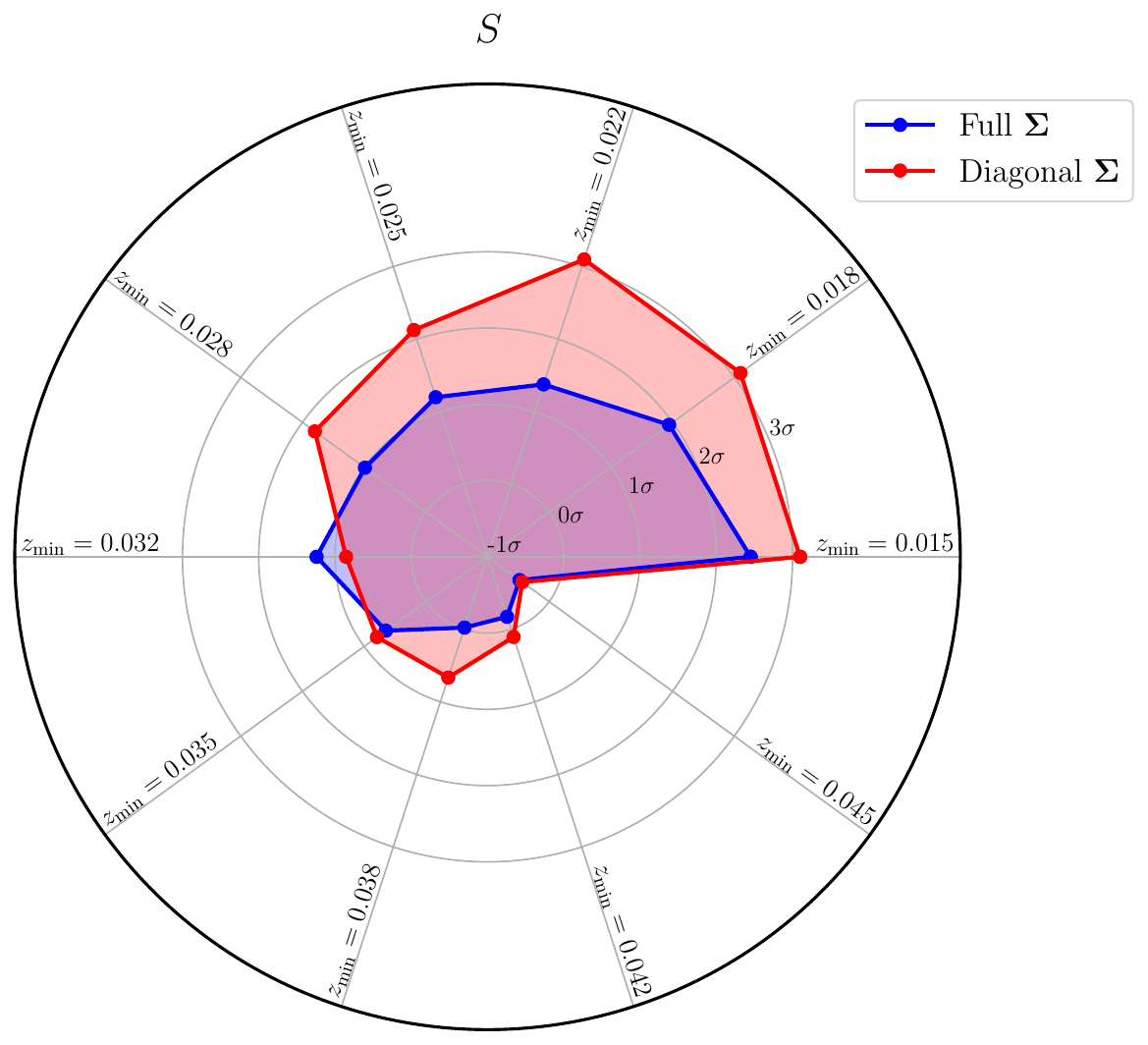}
\caption{The effect of the covariance matrix on the dipole amplitude
$A_{\rm dip}$ and the dipole significance estimator $S$
\eqref{eq:significance} as a function of $z_{\min}$. The diagonal covariance
matrix corresponds to restricting the $48\times48$ covariance matrix of the sky
patches to its diagonal elements only. The latter, as expected, generically underestimates the errors.}
\label{fig:fig_8}
\end{figure*}

\vspace*{-6mm}
\subsection{The dipole direction and its evolution with \texorpdfstring{$z_{\min}$}{zmin}}\label{sec:direction}
\vskip -3mm

In addition to the dipole amplitude, we have also assessed the direction of the dipole, as reported in Table \ref{tab:dipole-angle}. To visualize the results, we have depicted the inferred dipole directions in Fig.~\ref{fig:fig_42}. This figure shows that for $z_{\rm min} < 0.03$, the dipole directions cluster around a common region of the sky, remaining within approximately $30^\circ$ of the Shapley Supercluster. For the purpose of our directional analysis, the Shapley Supercluster region is defined around the central Galactic coordinates $(l, b) \approx (311.0^{\circ}, 30.8^{\circ})$ with an effective angular radius of $\sim 10^{\circ}$ (see, e.g., \cite{Proust:2005jt}). In fact, as one can see in Fig.~\ref{fig:fig_42}, the dipole is half-way between the Shapley and CMB dipole directions. For larger values of $z_{\rm min}$ ($>0.03$) the reconstructed dipole amplitude  drops and its direction gradually drifts away from the Shapley direction, as the dipole amplitudes also drops to zero. We caution the the reader again that for the three highest redshifts bins with $z_{\rm min} \geq  0.038$, the dipole amplitude is consistent with zero, so we should not assign any importance to directional shifts in the three highest redshift bins. Once these directions are removed, it is clear that the directional dependence of the dipole is robust to $z_{\rm min}$ changes where $A_{\rm dip}$ is non-zero.

{\begin{table*}[ht]
	\centering
	\begin{tabular}{|c||  c| c|  c| c|}
		\hline
		$z_{\min}$ &   $l$ & $b$ &
		$\theta_{\text{Shapley}}$ & $\theta_{\text{CMB}}$ \\
		\hline
		0.015 &   $307.5^\circ \pm 22.0^\circ$ & $61.2^\circ \pm 9.1^\circ$ & $30.5^\circ \pm 9.1^\circ$ & $27.6^\circ \pm 10.6^\circ$ \\
		0.018  & $298.6^\circ \pm 21.1^\circ$ & $52.5^\circ \pm 12.1^\circ$ & $23.5^\circ \pm 12.3^\circ$ & $22.2^\circ \pm 12.8^\circ$ \\
		0.022 &   $288.3^\circ \pm 27.1^\circ$ & $56.8^\circ \pm 14.1^\circ$ & $30.4^\circ \pm 14.4^\circ$ & $16.9^\circ \pm 14.7^\circ$ \\
		0.025 & $286.0^\circ \pm 30.4^\circ$ & $59.4^\circ \pm 15.7^\circ$ & $33.2^\circ \pm 15.6^\circ$ & $16.9^\circ \pm 15.5^\circ$ \\
		0.028 &  $289.8^\circ \pm 38.6^\circ$ & $65.3^\circ \pm 15.8^\circ$ & $36.9^\circ \pm 15.9^\circ$ & $21.8^\circ \pm 16.0^\circ$ \\
		0.032  & $284.0^\circ \pm 27.2^\circ$ & $59.4^\circ \pm 14.7^\circ$ & $33.9^\circ \pm 14.3^\circ$ & $16.1^\circ \pm 14.1^\circ$ \\
		0.035  & $296.3^\circ \pm 33.2^\circ$ & $55.1^\circ \pm 20.8^\circ$ & $26.5^\circ \pm 20.4^\circ$ & $20.9^\circ \pm 19.0^\circ$ \\
		0.038  & $356.1^\circ \pm 95.4^\circ$ & $70.8^\circ \pm 30.0^\circ$ & $46.9^\circ \pm 31.0^\circ$ & $45.8^\circ \pm 31.1^\circ$ \\
		0.042  & $20.6^\circ \pm 69.6^\circ$ & $60.7^\circ \pm 41.0^\circ$ & $53.6^\circ \pm 34.1^\circ$ & $59.7^\circ \pm 37.8^\circ$ \\
		0.045 & $228.5^\circ \pm 221.9^\circ$ & $77.0^\circ \pm 57.5^\circ$ & $58.4^\circ \pm 49.9^\circ$ & $31.9^\circ \pm 53.5^\circ$ \\
		\hline
	\end{tabular}
	\caption{{The $H_0$ dipole direction at each $z_{\min}$ and the angles between the dipole and Shapley  $\theta_{\text{Shapley}}$, and the CMB dipole $\theta_{\text{CMB}}$.}\label{tab:dipole-angle}}
\end{table*}}

\begin{figure}[ht!]
    \centering
    \includegraphics[width=\linewidth]{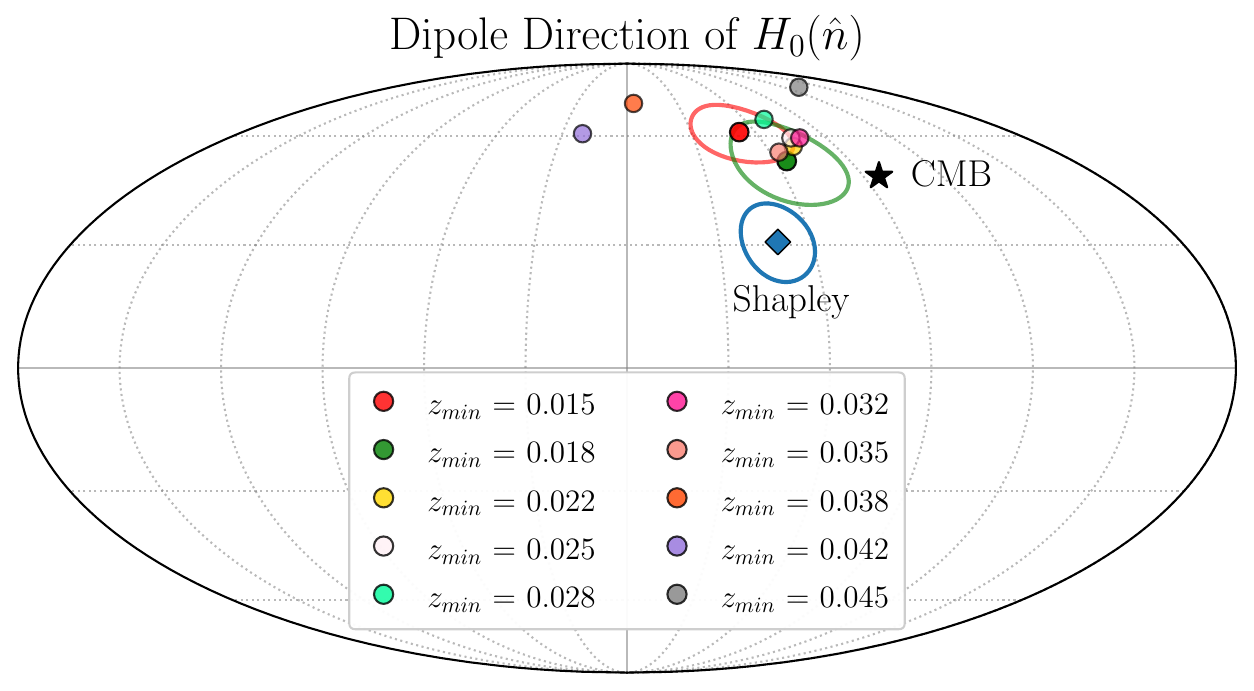}
    \caption{{Dipole direction as a function of $z_{\rm min}$. The directions of the CMB dipole and the Shapley Supercluster are provided for reference. The blue contour  around Shapley supercluster indicates its  spatial extent, while the red and green contours represent the positional uncertainties of  dipole directions of the first two bins corresponding to $z_{\rm min} = 0.015$ and $z_{\rm min} = 0.018$.} {We caution that beyond $z_{\rm min} = 0.038$, the dipole amplitude $A_{\rm dip}$ is effectively consistent with zero.}}
    \label{fig:fig_42}
\end{figure}

As Table~\ref{tab:dipole-angle} shows, our angular resolution power, especially for the low $z_{\min}$, while not so good, is good enough to show (1) alignment in the dipoles of the $z_{\min}<0.038$  bins (where the dipole amplitude is not negligible); (2) that they are in the sky region broadly consistent with both Shapley and the  CMB dipole direction. If the detected dipole were of cosmological origin, one would not expect the decreasing trend in the dipole amplitude with $z_{\min}$, and its direction would be expected to remain stably aligned with the CMB dipole. This is hinting  that the dipole is likely to be a local effect, and may be associated with the local velocity fields and bulk flows, compatible with the findings of cosmicflows-4 \cite{Valade:2024riw, Salzano:2025ang}.

\vspace*{-6mm}
\subsection{Anisotropic expansion beyond the dipole}\label{sec:internal-consistency}
\vskip -3mm
One can  quantify other angular information in $H_0(\hat n;z_{\text{min}})$ maps by a formal expansion of this function on the celestial sphere in terms of spherical harmonics. However, since higher {multipoles} beyond the dipole are expected not to have a significant contribution, we focus on a simpler measure as an indicator of beyond dipole information in the expansion rate $H_0(\hat n;z_{\text{min}})$. 

To this end, we compute the directional contrast across opposite sky patches,
\begin{equation}
    \Delta\text{H}_0^{(i)} = |H_0(\hat n_i) - H_0(-\hat n_i)|, 
\end{equation}
{which mirrors the indicator from \cite{Krishnan:2021jmh, Zhai:2022zif, McConville:2023xav}.}
Since the $H_0$ maps are constructed on a HEALPix grid with 48 pixels, each pixel has a corresponding antipodal counterpart, yielding 24 opposite pairs on the sky for each redshift bin denoted by $z_{\text{min}}$, and define {$\Delta \text{H}^{\rm max}_0$ for each $z_{\text{min}}$} as the largest value among these 24 contrasts for each redshift bin.

The motivation for this test is straightforward. If the dipole was the only contributor to the anisotropy in the expansion rate, as assumed in (\ref{eq:fit_dipole}), one has  
\begin{equation}
    H_0(\hat n)-H_0(-\hat n)=2\,\mathbf{D}\cdot\hat n,
\end{equation}
and its maximum value is expected to be $2|\mathbf{D}|=2A_{\text{dip}}$. So, if the anisotropy in $H_0$ values is dominated by a dipole, the quantity $\Delta\text{H}^{\text{max}}_0/2$ should be statistically consistent with the fitted dipole amplitude $A_{\text{dip}}$. However, since $\mathbf{D}\cdot\hat n\leq A_{\text{dip}}$ and moreover, in principle there could be higher pole contributions, $\Delta\text{H}^{\text{max}}_0/2 \geq A_{\text{dip}}$, as seen from Table \ref{tab:deltaH}. 

To quantify the agreement with (or the departure from) the dipole, we define the $\beta$ estimator,\footnote{We note that the estimator $$\frac{\Delta\text{H}^{\text{max}}_0/2-A_{\text{dip}}}
	{\sqrt{\sigma_{_{\Delta\text{H}^{\text{max}}_0/2}}^2+\sigma_{A_{\text{dip}}}^2}},
$$ with $\sigma_{A_{\text{dip}}}$ being the propagated uncertainties on $A_{\text{dip}}$, would underestimate $\beta$ because the errors on $\Delta\text{H}^{\text{max}}_0$ and $A_{\text{dip}}$ are not independent. The $\beta$ given in \eqref{beta-anisotropy-vs-dipole} {provides a conservative upper bound on the statistical significance of the difference between $\Delta\text{H}_0^{\rm max}/2$ and $A_{\rm dip}$.}}
\begin{equation}\label{beta-anisotropy-vs-dipole}
    \beta:=\frac{\Delta\text{H}^{\text{max}}_0/2-A_{\text{dip}}}
	{\sigma_{_{\Delta\text{H}^{\text{max}}_0/2}}},
\end{equation}
where $\sigma_{\Delta\text{H}_0/2}$ is the propagated uncertainties on $\Delta \text{H}_0/2$. We find that $\beta<1$ for all redshift cuts considered in this work. This demonstrates that the largest antipodal contrast in the reconstructed maps is fully consistent with the dipole amplitude inferred from the global fit, supporting the interpretation that the dominant angular variation in our directional $H_0$ maps is indeed dipolar.

	\begin{table}
		\centering
		\begin{tabular}{|c|| c| c| c| c|}
			\hline
			$z_{\min}$ 
            & $\quad{\Delta\text{H}^{\text{max}}_0}/{2}\quad$ & ${A_{\rm dip}}$  & $\quad\beta\quad$ \\
			\hline
			0.015 
            & $1.91 \pm 0.83$ {($2.3 \sigma$)} & $1.16 \pm 0.28$ &$0.33$ \\
			0.018 
            & $1.96 \pm 0.84$ {($2.3 \sigma$)} & $1.01 \pm 0.28$ & $0.41$ \\
			0.022 
            & $1.82 \pm 0.91$ {($2.0 \sigma$)} & $0.87 \pm 0.30$ & $0.48$ \\
			0.025 
            & $2.04 \pm 0.91$ {($2.2 \sigma$)} & $0.91 \pm 0.34$ & $0.51$ \\
			0.028 
            & $1.80 \pm 0.97$ {($1.9 \sigma$)} &  $0.93 \pm 0.39$ & $0.46$ \\
			0.032 
            & $1.57 \pm 1.02$ {($1.5 \sigma$)} & $1.05 \pm 0.39$ & $0.35$ \\
			0.035 
            & $1.52 \pm 0.97$ {($1.6 \sigma$)} & $0.87 \pm 0.43$ & $0.41$ \\
			0.038 
            & $1.82 \pm 1.04$ {($1.8 \sigma$)}& $0.56 \pm 0.46$ & $0.70$ \\
			0.042 
            & $1.40 \pm 1.08$ {($1.3 \sigma$)} & $0.49 \pm 0.45$ & $0.70$ \\
			0.045 
            & $1.36 \pm 1.01$ {($1.3 \sigma$)} & $0.35 \pm 0.52$ & $0.78$ \\
			\hline
		\end{tabular}
		\caption{Maximum antipodal $H_0$ contrasts $\Delta\text{H}^{\text{max}}_0(z_{\min}), {A_{\rm dip}}$   and $\beta$, cf. \eqref{beta-anisotropy-vs-dipole}. $\beta<1$ means anisotropy in the expansion rate is dipole dominated.}
		\label{tab:deltaH}
	\end{table}

We close this section with some comments on the statistical significance of the antipodal $\Delta\text{H}^{\text{max}}_0$ values in Table \ref{tab:deltaH}. Nothing prevents a statistically significant dipole characterized by the low $p$-values ($\gtrsim 3.1 \sigma$) for $A_{\rm dip}$, nevertheless it is more relevant that the dipole impacts a physical observable either the Hubble constant $H_0$ \cite{Krishnan:2021jmh, Zhai:2022zif, McConville:2023xav} or a peculiar velocity, e. g. \cite{Sorrenti:2022zat}. As noted in \cite{Krishnan:2020vaf}, $H_0$ has an interpretation in terms of an integration constant in the Friedmann equations, so mathematical consistency of the FLRW framework demands it is a constant. As a result, the null hypothesis is $\Delta\text{H}^{\text{max}}_0 = 0$. Furthermore, antipodal SNe are effectively independent exhibiting weak correlations, which means that $H_0$ determinations from antipodal non-overlapping SNe samples are essentially statistically independent.

The first observation is that at $z_{\rm min} = 0.015$ the statistical significance of $A_{\rm dip}$ based on mocks exceeds $3.1 \sigma$ but the  statistical significance of $\Delta\text{H}^{\text{max}}_0$ is considerably lower at $2. 3 \sigma$. A key consistency check of different estimators is that the statistical significance decreases as $z_{\rm min}$ increases. Nevertheless, it is interesting that the statistical significance of $\Delta\text{H}^{\text{max}}_0$ decreases to $1.3 \sigma$ at $z_{\rm min} = 0.045$, well beyond the redshift where $A_{\rm dip}$ becomes consistent with zero. Thus, even at deeper redshifts one may have a non-zero $\Delta\text{H}^{\text{max}}_0$ that is inconsistent with zero despite a zero dipole. There are basically two interpretations here: (A) The residual $ \langle \Delta\text{H}^{\text{max}}_0 \rangle \neq 0$ is a dipole but the data is not constraining enough and the dipole ansatz \eqref{eq:fit_dipole} loses sensitivity; (B) The dipole drops out because $ \langle \Delta\text{H}^{\text{max}}_0 \rangle \neq 0$ is only tracking higher order multipoles. Note, the problem pursuing these multipoles is that it necessitates extra parameters beyond a dipole, thus inflating errors further and making it less likely that one recovers a signal from the currently available data.

\vspace*{-4mm}
\section{Comparison with the existing literature} \label{sec:comparison}
\vspace*{-2mm}

So far we have observed that the  dipole in the expansion rate (both amplitude and direction) from Pantheon+ compilation, is pronounced at lower $z_{\min}$. However, as we go to $z_{\min} \geq 0.038$ the significance of the amplitude {drops to a value consistent with zero making further interpretation moot. Where $A_{\rm dip} > 0$, the direction of the dipole is robust to $z_{\rm min}$ cuts and remains close to the CMB dipole and Shapley supercluster directions with a marginal preference for the former. The fact that the statistical significance of all estimators ($p$-values, $S$, $\Delta\text{H}^{\text{max}}_0$) decreases with increasing $z_{\rm min}$  suggests that the dipole is a low-redshift feature which may be caused by large structures, potentially even the Shapley supercluster. The fact that we lose sensitivity to the dipole at $z_{\rm min} = 0.038$ suggests that the data to map out Shapley supercluster with SNe and a dipole ansatz is currently not available.}

\textbf{Comparison to \cite{Andrade:2017iam, Andrade:2018eta}:} The authors  use the earlier Pantheon SNe dataset \cite{Pan-STARRS1:2017jku} and conclude that the data are consistent with statistical isotropy. This conclusion seems to be in tension with our results. However, there are several important methodological differences which help explain the discrepancy. First, the Pantheon+ compilation employed in this work contains substantially more low $z$ SNe than the  Pantheon sample used in \cite{Andrade:2017iam, Andrade:2018eta}, providing considerably greater sensitivity to local physics (velocity fields and large-scale structures). As demonstrated by our tomographic analysis, the anisotropic signal is primarily driven by SNe at $z\lesssim0.03$, precisely the regime where the increased low-redshift statistics of Pantheon+ become most relevant. Second, the anisotropy estimator adopted in \cite{Andrade:2018eta} differs fundamentally from the one employed here. Their analysis is based on the difference between the maximum and minimum values of the reconstructed cosmological parameters across the sky, effectively comparing extrema of directional maps. While useful as a global anisotropy indicator, such a statistic is not specifically optimized to detect a coherent dipolar pattern. In contrast, our analysis reconstructs full-sky $H_0(\hat{n})$ maps and directly extracts the dipole component, providing a more targeted characterization of large-scale directional variations.

\textbf{Comparison to \cite{Lopes:2024vfz}:} Our results are more closely aligned with the findings of Lopes et al. who perform a directional analysis of 501 Pantheon+ SNe in the range $0.015\leq z\leq0.06$. Using overlapping spherical caps and directional measurements of the local Hubble expansion, a statistically significant dipole toward $(l,b)=(326.1^\circ\pm11.2^\circ, 27.8^\circ\pm11.2^\circ)$ is detected, which is remarkably close to the Shapley supercluster direction, which is slightly different than what we find here. The authors further interpreted this signal as evidence for a bulk flow velocity of $132.14\pm109.3~{\rm km/s}$, which is consistent with the general picture resulting from our analyses. Their methodology involves using a different mocking strategy, \textit{the shuffling SNe mocking}. In comparison, we use \textit{MC-lcdm} mocks, which as we argue, are more apt for providing isotropic reference configurations, yielding different $\sigma$-level significances. We use the same mocks for reading the covariance matrix and carefully propagate errors from Pantheon+ dataset to the $ 48\times 48$ covariance matrix. We have traced that using the covariance matrix or only its diagonal part can yield the difference in the dipole directions between our results and those in \cite{Lopes:2024vfz}.\footnote{Some aspects of error propagation in  \cite{Lopes:2024vfz} are puzzling. It is not clear how  the covariance matrix of Pantheon+ is dealt with in the shuffling SNe simulations adapted there. For example, in equation (9) of the paper an ad hoc dimensionless statistical error is added to a dimensionful $H_0$ error and the authors are unclear how the dipole angular power $C_1$ handles correlated $H_0$ values.}

{Our study extends the analysis of \cite{Lopes:2024vfz} in an important way, which is highlighted in the title of this paper: Their work focuses on a fixed redshift interval, while we perform  tomographic reconstruction allowing us to uncover the redshift  evolution of the dipole in the expansion rate. We find that the dipole amplitude decreases monotonically from $1.16\pm0.28~{\rm km/s/Mpc}$ at $z_{\rm min}=0.015$ to $0.35\pm0.52~{\rm km/s/Mpc}$ at $z_{\rm min}=0.045$, while its statistical significance falls from above $3\sigma$ to values consistent with isotropy. This redshift dependence {suggests} that the observed anisotropy is predominantly associated with nearby structures rather than a  cosmological effect \footnote{Note, different estimators have different sensitivities to any putative anisotropy, since $A_{\rm dip}$ becomes statistically equivalent with zero before $\Delta \text{H}_0^{\rm max}$ as $z_{\rm min}$ increases. Evidently, different estimators can disagree on the $z_{\rm min}$ at which one recovers an isotropic Universe.} {We have also employed $\beta$ to argue that beyond the dipole any contribution to expansion anisotropy is small, where it should be borne in mind that the errors are large, i. e. $\Delta\text{H}_0^{\rm max} \sim 2 \pm 1$ km/s/Mpc. This test will become better as SNe data improves.}

\textbf{Comparison to \cite{Sorrenti:2022zat, Sorrenti:2024ugq, Sorrenti:2024ztg}.}
These works analyze redshift and angular distribution of the local velocity field which can be interpreted as similar result for the expansion rate, once one recalls the celebrated Hubble law. {Most of the analysis is conducted with heliocentric redshifts $z_{\rm HEL}$, so one expects to recover the CMB dipole, while the authors find a statistically significant bulk flow that differs largely in direction, not magnitude, that precludes this happening. Given the difference in redshifts, the null hypothesis differs; in $z_{\rm CMB}$ one expects no dipole, whereas in $z_{\rm HEL}$ one expects the CMB dipole. A key difference is that \cite{Sorrenti:2022zat} (see also \cite{Horstmann:2021jjg}) never recovers the expected null hypothesis, and if true, this raises questions about SNe samples, especially at the lowest redshifts.} 

{Instead, \cite{Sorrenti:2022zat} reports evidence for a coherent bulk velocity of approximately $350~{\rm km/s}$ extending to $z\approx0.0375$. They further show that the significance of the signal decreases as increasingly nearby SNe are removed from the sample. This behavior closely mirrors the trend identified in our more systematic tomographic analysis, where the dipole significance rapidly declines once SNe with $z\lesssim0.038$ are excluded. Additional indirect support for a local origin of the signal comes from the cosmographic analysis of~\cite{Sorrenti:2024ugq} which detects a significant radial infall velocity of approximately $100~{\rm km/s}$ within the nearby Universe focusing on the monopole local velocity fields rather than directional $H_0$ maps. Furthermore, the subsequent multipole analysis of ~\cite{Sorrenti:2024ztg} reveals significant monopole and quadrupole contributions in addition to the dipole, suggesting that the local velocity field may possess a rich redshift and angular structure than can be captured by a simple isotropic pure bulk-flow model.}

{\textbf{Comparison to \cite{Krishnan:2021jmh, Zhai:2022zif, McConville:2023xav}.}} {These papers consider hemisphere decompositions of SNe with estimator $\Delta{\text{H}}_0^{\rm max}$ both with an without Cepheids, with $z_{\rm HD}$ redshifts corrected for both the CMB dipole and SNe peculiar velocities, namely SNe redshifts used to determine $H_0$. (Here we used $z_{\text{CMB}}.$) Ref. \cite{Krishnan:2021jmh} studies Pantheon SNe,  \cite{Zhai:2022zif} studies Pantheon combined with Cepheids, while \cite{McConville:2023xav} studies Pantheon+, with Cepheids  incorporated. Overall, the statistical significance of $\Delta\text{H}_0^{\rm max}$ is $\lesssim 2 \sigma$, but a direction close to the CMB dipole direction is tracked by $\Delta\text{H}_0^{\rm max}$. A key difference is that \cite{Zhai:2022zif, McConville:2023xav} allow $M$ to vary by splitting the Cepheids on the sky and note that calibration has a bearing on the $H_0$ dipole. Refs.~\cite{Krishnan:2021jmh, McConville:2023xav} consider removing lower redshift SNe, motivated in part by Shapley \cite{McConville:2023xav}, to understand how $\Delta\text{H}_0^{\rm max}$ degrades. Modulo the differences in redshift and calibration, \cite{McConville:2023xav} reports a $1.9 \sigma$ $\Delta\text{H}_0^{\rm max}$, which agrees well with the $2.2 \sigma$ $A_{\rm dip}$ and $2.0 \sigma$ $\Delta{\text{H}}_0^{\rm max}$ at $z_{\rm min} = 0.022$. Given the overlap with the SH0ES collaboration \cite{Riess:2021jrx}, \cite{McConville:2023xav} is largely concerned with whether $H_0$ can be determined to $1 \%$. The latter is obviously of great interest and will be the main point we will be discussing in the third part of the trilogy of our papers.}

\vspace*{-4mm}
\section{Discussion} \label{sec:Discussion}
\vspace*{-2mm}

There is no shortage of overlapping and conflicting reports of anisotropies in the literature. The anisotropies are more often than not traced by different cosmological parameters ($H_0, \Omega_m, q_0$,\dots) that are sensitive to different redshift ranges. The literature also makes use of a host of different indicators. For these reasons, it is unsurprising that findings often disagree.

{Nevertheless, if one tacks close to physical expectations there is usually a grain of truth. The cosmicflows program \cite{Courtois:2011fa, Tully:2013wqa, Tully:2016ppz, Tully:2022rbj} has mapped out galaxy peculiar velocities in the local universe, with samples much larger than SNe samples, revealing fascinating local structure. Of key interest is the observation that the local Universe is anisotropic {at least to $\sim 400$Mpc distances} due to a bulk flow of galaxies in a direction consistent with the Shapley supercluster \cite{Hoffman:2017ako} - close to the CMB dipole direction. Through recent data upgrades, the statistical significance of the bulk flow has increased to the point that it may challenge $\Lambda$CDM \cite{Watkins:2023rll, Whitford:2023oww}. A statistically significant local bulk flow is expected to modulate $H_0$ on the sky. Tellingly, higher $H_0$ values in the direction of the CMB dipole have been independently reported since 2021 in SNe samples \cite{Krishnan:2021jmh, Zhai:2022zif, Kalbouneh:2022tfw, McConville:2023xav, Perivolaropoulos:2023tdt, Hu:2023eyf, Lopes:2024vfz, Sah:2024csa}, admittedly at low statistical significance (see also \cite{Krishnan:2021dyb, Luongo:2021nqh} for other observables) and also in cosmicflows-4 (CF4) \cite{Boubel:2024cmh, Courtois:2025xcs}. Earlier observations point to dipoles in different directions \cite{McClure:2007vv, Schwarz:2007wf} (see also \cite{Migkas:2020fza, Migkas:2021zdo}), but this more recent synergy between a bulk flow in cosmicflows and a $H_0$ modulation suggests a physical origin.}

To increase our confidence in the synergy, it is imperative to revisit the feature with different methods to test robustness. Here, we performed a fit of a dipole ansatz to a sky-map of $H_0$ determinations from overlapping $75^{\circ}$ spherical caps in the Pantheon+ sample and  track SNe  into where one expects to be in the Hubble flow and fit matter density $\Omega_m$ to allow cosmological dependence. We carefully considered the error propagation by explicitly constructing a covariance matrix to handle correlated $H_0$ values due to overlapping SNe. A further novelty comes from the introduction of a low redshift cutoff $z_{\rm min}$ in a bid to see if the $H_0$ dipole is purely a low redshift feature or not.

Concretely, we considered three estimators: i) the $p$-value of the dipole amplitude $A_{\rm dip}$ with respect to isotropic $\Lambda$CDM mocks, ii) a conservative estimator $S$ that propagates the extra error in $A_{\rm dip}$ and iii) the maximal antipodal $H_0$ variation $\Delta\text{H}_0^{\rm max}$ with $\beta$ estimator. With our lowest $z_{\rm min}$ values, we recover $\sim 3 \sigma$ result also reported in \cite{Lopes:2024vfz}, and confirm that as $z_{\rm min}$ increases the statistical significance of all estimators decrease, which is what one expects if anisotropies are due to local structures. However, we find that while $A_{\rm dip}$ is essentially consistent with zero in the three highest redshift bins,  $\Delta\text{H}_0^{\rm max}$ still reports a non-zero result across all bins. One interpretation of this result is that while the anisotropy is well modeled by a dipole at lower redshifts (see $\beta$ indicator), it is plausible that higher multipoles explain the $\Delta\text{H}_0^{\rm max}$ at higher redshifts. Given that $A_{\rm dip}$ becomes consistent with zero at $z_{\rm min} = 0.038$ and its direction more marginally favors the CMB dipole, we are unable to trace the anisotropy to Shapley supercluster $0.03 \lesssim z \lesssim 0.06$ \cite{Quintana:2000vb}. Other methods may work but they may require an ansatz based on physical expectations \cite{Colin:2010ds}.} 

Our analysis makes use of $z_{\rm CMB}$, but in principle one should further correct SNe redshifts for peculiar velocities to work with $z_{\rm HD}$ that one usually uses to determine $H_0$. It will be interesting to see what difference this makes. However, we find good agreement with other studies that use $z_{\rm HD}$ where redshift ranges overlap the most e. g. \cite{McConville:2023xav}; this suggests that conventional peculiar velocity modeling will not make a big difference. Note, conventional models are unlikely to accommodate an anomalous bulk flow, e. g. \cite{Watkins:2023rll}. So, the dipole we find may still be attributed to/correlated with the bulk flow.

Finally, we close with two comments on the implications our findings for the CP and $H_0$ tension. Regarding the former, we note that observationally the CP is a statistical statement which is expected to be true beyond a distance or redshift $z_{\text{CP}}$, $z\gtrsim z_{\text{CP}}$, where $z_{\text{CP}}$ is where one is expanding with the Hubble flow \cite{Aluri:2022hzs}.  While traditionally one takes $ \gtrsim 100$ Mpc as the CP scale, recent observations have point to $\gtrsim 400$ Mpc \cite{Courtois:2025xcs}, scales at which superclusters are not yet in the point-particle regime.} So, our findings confirm these results but say nothing about the CP at {larger redshifts}. Nonetheless, our results can have important consequences for the $H_0$ tension, recalling that $H_0$, by the very definition, is a local quantity in principle measured at redshifts $z \lesssim 0.1-0.2$. If expansion rate varies by $\sim 10\%$ over the sky, then $1\%$ precision on determination of $H_0$ is ill-defined \cite{McConville:2023xav}.

\begin{acknowledgments}
The authors acknowledge the computational resources provided by the Scalable Analytics and Research Virtual Environment (SARVe) facility at the School of Theoretical Physics, IPM. MHJ-K acknowledges  Iranian National Science Foundation (INSF) postdoctoral fellowship grant No. 4040049. MMSh-J and SP would like to thank the support through INSF research chair grant No. 4045163. This article/publication is based upon work from COST Action CA21136– “Addressing observational tensions in cosmology with systematics and fundamental physics (Cosmo
Verse)”, supported by COST (European Cooperation in Science and Technology). 
\end{acknowledgments}

\appendix
\section{Analytical approximation for the covariance of the directional \texorpdfstring{$H_0$}{H0} estimates}
\label{app:covariance}
{{
In the main analysis presented in this work, the covariance matrix of the reconstructed directional Hubble parameters, $\boldsymbol{\Sigma}$, is estimated numerically using parametric bootstrap realizations, based on \textit{MC-lcdm} simulations. Nevertheless, it is instructive to derive an approximate analytical expression that explicitly relates the covariance of the inferred $H_0$ values to the underlying Pantheon+ covariance matrix. This appendix summarizes that derivation.

The apparent magnitude \eqref{eq:eq_2} can be written as
\begin{equation}
m_{\rm th}
=
25-\frac{1}{k}\ln\!\left(\frac{\mathcal H_0}{c}\right)
+\frac{1}{k}\ln\!\left[(1+z)d_L(z)\right],
\label{eq:app_mth}
\end{equation}
where
\begin{equation}
\mathcal H_0
=
H_0\,e^{-kM},
\qquad
k=\frac{\ln10}{5}.
\label{eq:app_calH}
\end{equation}
This parameterization makes explicit the well-known degeneracy between the Hubble constant $H_0$ and the SNe absolute magnitude $M$, showing that the observable depends only on the combination $\mathcal H_0$.

Assuming Gaussian likelihoods, minimizing $\chi^2_{\rm tot}(\hat n_i)$ \eqref{eq:chi2tot} leads to the maximum-likelihood estimators
\begin{equation}
\begin{split}
\ln\mathcal H_0(\hat n_i)
=
-k
\sum_a
w^a(\hat n_i)\,
m_a
+
\sigma_M^2,\\
       {\Omega}_m(n_i){}_{\text{\tiny{best fit}}}= k \sum_a w^a(\hat{n}_i)\ f_a m_a, 
\end{split}
\label{eq:app_lnH-Omega}
\end{equation}
where 
\begin{equation}
\begin{split}
 			w^a(\hat{n}_i) = \frac{\sum_b \big(\mathbf{C}^{\hat{n}_i}_{\rm SN}\big)^{-1}_{ab}}{\sum_{c,d} \big(\mathbf{C}^{\hat{n}_i}_{\rm SN}\big)^{-1}_{cd}},\\
			\sigma^{-2}_M = \sum_{ab} \big(\mathbf{C}_{\rm Ceph}\big)^{-1}_{ab}\\
            f_a=\frac{1}{d_L}\left(\frac{\partial d_{L}}{\partial {\Omega}_m}\right)_{z=z_a} 
\end{split}
			\label{eq:wstandard}       
\end{equation} }
Eq.~(\ref{eq:app_lnH-Omega}) shows that the reconstructed value of $\ln\mathcal H_0$ is simply a weighted linear combination of the observed SN magnitudes. Consequently, the covariance between two directional estimates follows directly from standard linear error propagation.
Using
\begin{equation}
\delta\ln\mathcal H_0(\hat n_i)
=
-k
\sum_a
w^a(\hat n_i)\,
\delta m_a,
\end{equation}
together with
\begin{equation}
{\rm Cov}(m_a,m_b)
=
\left(
\mathbf C_{\rm SN}^{\hat n_i\hat n_j}
\right)_{ab},
\end{equation}
one obtains
\begin{equation}
{\rm Cov}
\!\left[
\ln\mathcal H_0(\hat n_i),
\ln\mathcal H_0(\hat n_j)
\right]
=
k^2
\sum_{a,b}
w^a(\hat n_i)
\left(
\mathbf C_{\rm SN}^{\hat n_i\hat n_j}
\right)_{ab}
w^b(\hat n_j).
\label{eq:app_covlog}
\end{equation}
Finally, propagating from $\ln H_0$ to $H_0$ gives
\begin{equation}
\boxed{
\Sigma_{ij}
=
k^2
H_0(\hat n_i)
H_0(\hat n_j)
\sum_{a,b}
w^a(\hat n_i)
\left(
\mathbf C_{\rm SN}^{\hat n_i\hat n_j}
\right)_{ab}
w^b(\hat n_j)
}
\label{eq:app_covH0}
\end{equation}
where $\mathbf C_{\rm SN}^{\hat n_i\hat n_j}$ denotes the cross-block of the full Pantheon+ covariance matrix connecting the supernovae belonging to the two sky patches. Eq.~(\ref{eq:app_covH0}) provides an analytical approximation for the covariance matrix of the reconstructed directional $H_0$.  

Given the covariance matrix $\boldsymbol{\Sigma}$, the covariance of the fitted dipole parameters follows from the standard GLS result,
\begin{equation}
{\mathtt{C}}_{\boldsymbol{\theta}}
=
\left(
\mathbf{X}^{\rm T}
\boldsymbol{\Sigma}^{-1}
\mathbf{X}
\right)^{-1}.
\label{eq:Ctheta}
\end{equation}
The uncertainties of the derived dipole quantities $(A_{\rm dip},l,b)$ are then obtained through standard error propagation. The variance of the dipole amplitude is
\begin{equation}
\sigma_A^2
=
\frac{1}{A_{\rm dip}^2}
\mathbf{D}^{\rm T}
{\mathtt C}_D
\mathbf{D},
\label{eq:sigA}
\end{equation}
where ${\mathtt C}_D$ is the $3\times3$ sub-block of ${\mathtt C}_{\boldsymbol{\theta}}$ corresponding to $(D_x,D_y,D_z)$. The uncertainties in Galactic longitude and latitude follow from the Jacobian of the transformation in (\ref{eq:lb}),
\begin{equation}
\sigma_l^2
=
\frac{
D_y^2{\mathtt C}_{D,xx}
-
2D_xD_y{\mathtt C}_{D,xy}
+
D_x^2{\mathtt C}_{D,yy}
}
{(D_x^2+D_y^2)^2},
\label{eq:sigl}
\end{equation}
and
\begin{equation}
\begin{split}
\sigma_b^2
&=
\frac{1}{A_{\rm dip}^4}
\Big[
(D_x^2+D_y^2){\mathtt C}_{D,zz}
-
2D_z(D_x{\mathtt C}_{D,xz}+D_y{\mathtt C}_{D,yz})\\
&
+
\frac{D_z^2}{D_x^2+D_y^2}
\left(
D_x^2{\mathtt C}_{D,xx}
+
D_y^2{\mathtt C}_{D,yy}
+
2D_xD_y{\mathtt C}_{D,xy}
\right)
\Big].
\end{split}
\label{eq:sigb}
\end{equation}
These expressions fully propagate the covariance of the fitted GLS parameters into the uncertainties of the dipole amplitude and direction, which are used throughout the remainder of this work.
}

Although this expression offers useful physical insight into the origin of the inter-patch correlations, throughout this work we use the covariance matrix obtained directly from Monte Carlo realizations of the complete reconstruction pipeline, since that approach naturally captures all numerical effects of the inference procedure without relying on the linear approximation adopted above.


\begingroup\raggedright
\endgroup

\end{document}